\documentclass[aps,prb,twocolumn,superscriptaddress,showpacs]{revtex4-1}

\usepackage[utf8]{inputenc}
\usepackage[english]{babel}
\usepackage{graphicx}
\usepackage{amsmath,amsthm,amssymb,amsfonts}
\usepackage{xcolor}
\usepackage{bbold}
\usepackage{cancel}
\usepackage[colorlinks]{hyperref}

\definecolor{redred}{HTML}{D53E4F}
\newcommand{\Id}{{\mathbb 1}}

\begin{document}

\title{Dynamical Ginzburg criterion for the quantum--classical crossover of the Kibble--Zurek mechanism}

\author{Matthias Gerster}
\affiliation{Institute for Complex Quantum Systems \& Center for Integrated Quantum Science and Technologies, Universit\"at Ulm, D-89069 Ulm, Germany.}

\author{Benedikt Haggenmiller}
\affiliation{Institute for Complex Quantum Systems \& Center for Integrated Quantum Science and Technologies, Universit\"at Ulm, D-89069 Ulm, Germany.}

\author{Ferdinand Tschirsich}
\affiliation{Institute for Complex Quantum Systems \& Center for Integrated Quantum Science and Technologies, Universit\"at Ulm, D-89069 Ulm, Germany.}

\author{Pietro Silvi }
\affiliation{Center for Quantum Physics, Faculty of Mathematics, Computer Science and Physics, University of Innsbruck \& Institute for Quantum Optics and Quantum Information, Austrian Academy of Sciences, A-6020 Innsbruck, Austria.}
\affiliation{Institute for Complex Quantum Systems \& Center for Integrated Quantum Science and Technologies, Universit\"at Ulm, D-89069 Ulm, Germany.}

\author{Simone Montangero}
\affiliation{Institute for Complex Quantum Systems \& Center for Integrated Quantum Science and Technologies, Universit\"at Ulm, D-89069 Ulm, Germany.}
\affiliation{Theoretische Physik, Universit\"at des Saarlandes, D-66123 Saarbr\"ucken, Germany.}
\affiliation{Dipartimento di Fisica e Astronomia ``G. Galilei", Universit\'a degli Studi di Padova, I-35131 Italy.}

\date{\today}

\begin{abstract}
We introduce a simple criterion for lattice models to predict quantitatively the crossover between the classical and the quantum scaling of the Kibble--Zurek mechanism, as the one observed in a quantum $\phi^4$-model on a 1D lattice~[\href{https://doi.org/10.1103/PhysRevLett.116.225701}{Phys.\ Rev.\ Lett.\ \textbf{116}, 225701 (2016)}]. We corroborate that the crossover is a general feature of critical models on a lattice, by 
testing our paradigm on the quantum Ising model in transverse field for arbitrary spin-$s$ ($s \geq 1/2$) in one spatial dimension. By means of tensor network methods, we fully characterize the equilibrium properties of this model, and locate the quantum critical regions via our dynamical Ginzburg criterion. We numerically simulate the Kibble--Zurek quench dynamics and show the validity of our picture, also according to finite-time scaling analysis.
\end{abstract}


\maketitle

\section{Introduction}

Understanding the behavior of correlated matter when a physical system is driven out of equilibrium is a problem of paramount importance in classical and quantum mechanics, material science, and engineering.
In particular, the Kibble--Zurek (KZ) mechanism, the description of quasi-adiabatic quenches across a phase transition, has been studied both in classical and quantum scenarios,
spanning lengthscales from atomic sizes to galaxies~\cite{Kibble76,Zurek85,Kibbleagain1980,Dziarmaga98,Chuang91,Ducci99,IsingDQPT,IsingDQPT2,IsingDQPT3,hadzibacic:2015,Polkovnikov:2005,Sondhi:2012,degrandi:2011,Polturak:2003,KZ2016Labeyrie,KZ2017Vishwa,KZ2017Berdanier,KZ2017Meier,KZ2018Kennes}.
With the advent of quantum technologies --- enabled by recent advancements in experimental platforms based on atomic, molecular and optical physics ---
the KZ mechanism keeps being practical as well as fundamental. Indeed, quasi-adiabatic or beyond-adiabatic\cite{IsingDQPT3,Adolfo2014} quenches are still the most straightforward method for realizing complex quantum phases of matter in real experiments and to perform adiabatic quantum computations, {\it e.g.}, quantum annealing to solve classical hard problems~\cite{Georgescu2014,Lucas2014}.
Similarly, from a theoretical perspective, the KZ framework is a key scenario to deeply understand the interface between the classical macroscopic and the quantum microscopic world, especially in the context of critical phenomena and phase transitions, where the two worlds display quantitatively and qualitatively different emergent collective behaviors.

One particular example of the interplay, or rather competition, between the classical and the quantum KZ mechanism was recently numerically observed in Ref.~\onlinecite{PsiKZCrossover2016}, by some of the authors, in quenches across the linear-zigzag phase transition of ion coulomb crystals. They showed that two distinct regimes of quench times $\tau_Q$ emerge: A slow regime where the scaling of defects with~$\tau_Q$ is governed by a quantum theoretical description, and a fast regime where the defects scale according to a mean-field theory prediction, equivalent to a classical (zero-temperature) phase transition treatment. The crossover timescale between these two regimes (classical and quantum) can be roughly estimated by means of the Ginzburg criterion~\cite{Amit}, {\it i.e.\@} by comparing the order parameter with its own fluctuations.

In this work, we argue that such a crossover is not limited to a specific model: We show that this effect appears in the paradigmatic example for second order quantum phase transitions --- the one-dimensional Ising model in transverse field --- for any spin representation~$s$. 
We first fully characterize the phase diagram, and then analyze the KZ mechanism of the model focusing on the quantum--classical crossover for $\frac{1}{2} \leq s \leq 5$. As the Ginzburg criterion delivers imprecise quantitative predictions for $s \gg 1/2$ (see Appendix~\ref{sec:comparison-criteria}),
we propose a simple argument based on the properties at equilibrium, the Dynamical Ginzburg Criterion (DGC),
to better predict at which quench times $\tau_Q^{\times}$ the crossover is expected to occur in lattice models.
This prediction is practical and quantitative, allowing an arbitrary experimental platform to quickly test whether the crossover timescales are reachable within the platform specifications and typical coherence times.


\section{The Kibble--Zurek argument}

The KZ picture predicts a scaling law of the density of defects $n$ during a linear quench across a phase transition, as a function of the quench rate
(or the total quench time~$\tau_Q$) \cite{Kibble76,Zurek85}.
It is based on the assumption that at quasi-equilibrium the system has a response timescale $\tau_R(t)$ which
scales as $\tau_R \propto |h-h_c|^{-\nu z}$ with the distance from the critical point~$h_c$ of the driving parameter
\begin{equation}
	\label{eq:driving-param}
	h(t) = h_c + t \cdot \Delta h / \tau_Q \; ,
\end{equation}
controlling the Hamiltonian $H(h)$.
During the quench, the system follows the adiabatic trajectory as long as the relaxation time $\tau_R$ is shorter than the driving timescale~$\tau_D$, that is, the inverse relative rate of change of any scaling quantity $q$ of the system: $\tau_D = |q/\dot{q}|$. For a linear ramp quench, we thus obtain $\tau_D \propto |t|$.
As the system response slows down, we encounter a specific instant~$\hat{t}$ (freeze-out time) when the system abandons the adiabatic trajectory: The dynamics of the order parameter thus
freezes out, and the density of defects $n$ in the order is given by the equilibrium correlation length $\xi$ at this instant, $\hat{n} = \xi^{-1}(h(\hat{t})) \propto |h(\hat{t})-h_c|^{\nu}$.
This occurs when the response becomes slower than the driving, {\it i.e.\@} when $\tau_D(\hat{t}) \simeq \tau_R(\hat{t})$.
Combining all scaling laws delivers $\hat{t} \propto \tau_Q^{\nu z / (1+\nu z)}$, or equivalently $| \hat{h} - h_c | \propto \tau_Q^{-1 / (1+\nu z)}$ with
$\hat{h} = h(\hat{t})$,
and in turn $\hat{n} \propto \tau_Q^{- \kappa}$ with the KZ exponent $\kappa = \nu / (1+\nu z)$.
In this expression, $\nu$ and $\nu z$ are the scaling exponents of lengthscales and timescales, respectively, 
and they depend on whether the order parameter is ruled by a classical- or quantum critical scaling.

\subsection{The quantum scaling}

In the quantum regime, outside of a quantum critical point, the energy gap remains finite and directly determines the relaxation timescale~\cite{Damskisimplest,IsingDQPT,IsingDQPT2}.
Precisely, at quasi-equilibrium where the system occupies mostly the ground state of the instantaneous Hamiltonian $H(t)$, the slowest response timescale of the system is given by $\tau_R \simeq \hbar / E_\text{gap}(t)$, 
where $E_\text{gap}$ is the energy difference between the first excited state and the ground state of $H$. $E_\text{gap}$ is an equilibrium property, and near the critical point it scales with the control parameter, {\it i.e.\@} the external field $h$, as $E_{\text{gap}} = \varphi |h - h_c|^{\nu z}$. Equivalently, it scales with the correlation length of the order parameter as $E_{\text{gap}} \propto \xi^{-z}$. Consequently, the relaxation timescale~$\tau_R$ scales with the critical exponents~$\{z,\nu\}$ from the quantum critical point at equilibrium, which can be extracted by the corresponding conformal field theory based on dimensionality and symmetry breaking. For the universality class of the quantum Ising model in one spatial dimension, these exponents are $\nu = z = 1$, regardless of the local spin representation~$s$, as verified numerically in Appendix~\ref{sec:equilibrium-properties-spinS-Ising}.
The KZ exponent of the quantum regime is thus $\kappa = 1/2$.

\subsection{The classical scaling}

Conversely, in the mean-field (or classical) regime, the scaling exponents of the relaxation timescale~$\tau_R$ follow from the effective time-dependent Ginzburg equation for the order parameter $\phi$~\cite{Lagunaonequarter,Lagunaonethird}. Specifically, by requesting that the Ginzburg equation scales covariantly, we are able to identify the corresponding scaling exponents for $\tau_R$ and $\xi$ with respect to $h' = h-h_c$. The Ginzburg equation for a model with Ising criticality, a $Z_2$ symmetry breaking, is the one obtained from the Lagrangian of the $\phi^{4}$-model and reads~\cite{Lagunaonethird}
\begin{equation}
	\partial^2_t \phi - \partial^2_x \phi + h' \phi + \phi^3 = 0 \; ,
\end{equation}
where we consider both noise and damping to be negligible.
We now perform the scale transformation
\begin{equation}
	h' \to \lambda h'\, , \; 
	\phi \to \lambda^{\beta} \phi\, , \;
	x \to \lambda^{-\nu} x\, , \;
	t \to \lambda^{-\nu z} t \, ,
\end{equation}
and require covariance of the Ginzburg equation. This delivers $\nu = 1/2$ and $z = 1$ (as well as $\beta = 1/2$).
The KZ exponent of the classical regime is therefore $\kappa = 1/3$, quantitatively different from the quantum case.

\section{The Dynamical Ginzburg Criterion} \label{sec:DGC}

We adopt the following criterion to predict whether around a given quench time~$\tau_Q$ we expect to see the quantum or the classical scaling: We first estimate quantitatively the correlation length at equilibrium $\hat{\xi} = \xi(\hat{h})$ at the freeze-out point~$\hat{h}$ for that specific quench time~$\tau_Q$. If this correlation length is larger than the lattice spacing~$a$ ($\hat{\xi}(\tau_Q) \gg a$), then we expect to observe the quantum KZ scaling. Conversely, if it is smaller ($\hat{\xi}(\tau_Q) \ll a$) we expect to see the classical KZ scaling. We motivate this criterion based on the following argument: Consider a quantum system where the correlation length $\xi$  for some order parameter is smaller than the lattice constant. Then, the properties of such order are not ruled by entanglement, but only by local quantities.
If the entanglement does not play a role, then the mean-field picture is a reliable description for this type of order. Therefore, during the quench, if the system is not given sufficient time to build up quantum correlations leading to a $\hat{\xi}$ larger than the lattice constant, then, at freeze-out, the mean-field description of the order is still valid: We expect to observe the classical KZ scaling resulting from the scaling exponents of the mean-field (Ginzburg) picture. Conversely, if the quench times~$\tau_Q$ are sufficiently large so that~$\hat{\xi}$ is larger than~$a$, then the order properties at freeze-out are ruled by entanglement, thus the quantum KZ scaling will emerge. 

\begin{figure}
	\includegraphics[width=\columnwidth]{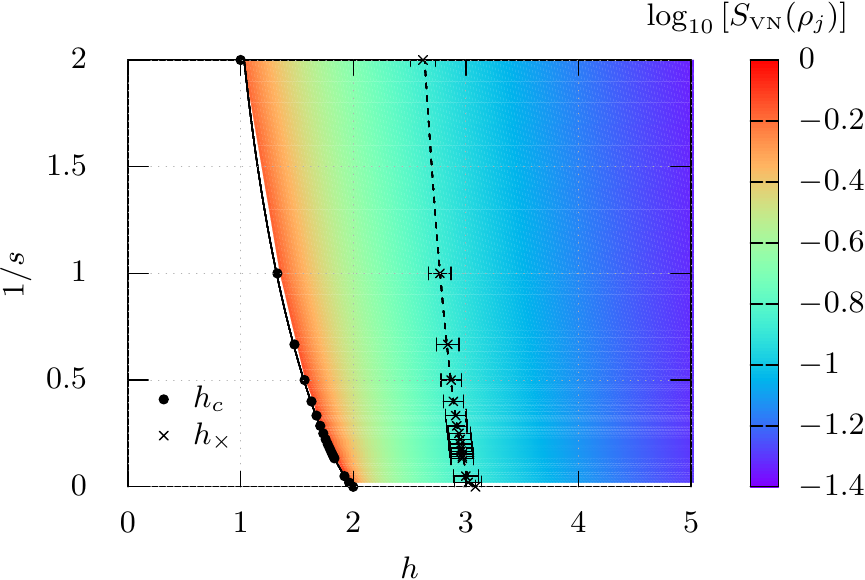}
	\caption{\label{fig:phasediagram}
		Phase diagram of the quantum Ising model~\eqref{eq:Isingclass} in 1D as a function of the external field $h$ and the inverse spin $1/s$. The black dots represent the phase transition points at the thermodynamical limit estimated via the von Neumann entropy (see Appendix~\ref{sec:criticalpoints}), while the solid black curve crossing them shows the fit $h_c(s) = h_c(\infty) - \tilde{a} ( \left| \ln s \right| - \tilde{b} ) / s$.
		These points separate the ferromagnetic phase, with white background, from the paramagnetic phase, with colored background. The crosses with error bars show the DGC points $h_\times$, estimated numerically via DMRG. The dashed line is a power-law fit in $1/s$ of the deviation $h_\times(\infty) - h_\times(s)$. The color in the paramagnetic phase encodes the single-site entanglement entropy~$S_\mathrm{VN}(\rho_j)$.
	}
\end{figure}

To make this argument quantitative, we start by estimating the dynamical quantum critical region, {\it i.e.\@} the value of external field $h_{\times}$ at which $\hat{\xi}(h_{\times}) = a$ at equilibrium, which lies in the disordered phase (see Fig.~\ref{fig:phasediagram}).
We perform this estimation via numerical simulations at equilibrium. Then, we exploit $\tau_R \simeq \hbar / E_\text{gap}$ and $E_\text{gap} \simeq \varphi |h-h_c|^{\nu z}$, where the scaling prefactor $\varphi$ is  calculated numerically. For estimating the driving timescale $\tau_D$ we adopt $\tau_D \simeq |\epsilon(t) / \dot{\epsilon}(t)| = |t|$, where $\epsilon(t) = h(t) - h_c$~\cite{IsingDQPT}. Under these assumptions the KZ equation $\tau_D(\hat{t}) = \tau_R(\hat{t})$ becomes
\begin{equation}
	\label{eq:tHat1}
	\hat{t} = \hbar \, |\hat{h}-h_c|^{-\nu z} / \varphi \; .
\end{equation}
Using the definition of the driving parameter $h(t)$ from Eq.~\eqref{eq:driving-param}, the freeze-out time can also be expressed as
\begin{equation}
	\label{eq:tHat2}
	\hat{t}=(\hat{h}-h_c) \, \tau_Q/\Delta h \; .
\end{equation}
Combining Eqs.~\eqref{eq:tHat1} and~\eqref{eq:tHat2} yields
\begin{equation}
	\tau_Q = \frac{\hbar \cdot |\Delta h|}{\varphi \cdot |\hat{h}-h_c|^{1+\nu z}} \; ,
\end{equation}
which allows to quantify the crossover quench time as
\begin{equation}
 \tau_Q^{\times} = \frac{\hbar \cdot |\Delta h|}{\varphi \cdot |h_\times - h_c|^{1+\nu z}} \; ,
 \label{eq:taucross}
\end{equation}
discriminating timescale regimes where the quantum ($\tau_Q \gg \tau_Q^{\times}$) or the classical ($\tau_Q \ll \tau_Q^{\times}$) KZ scaling will respectively emerge.
As an additional requirement to actually observe the classical KZ scaling, the quench must start outside the dynamical quantum critical region or the mean-field description will never be valid: This translates to a condition on the parametric quench interval, which reads
$|\Delta h| \gg |h_{\times} - h_c|$.
The parametric DGC point~$h_{\times}$ is thus a relevant point in the phase diagram, representing where the correlation length is equal to the lattice spacing, at equilibrium in the disordered phase.

\section{Numerical results}

In the following, we discuss numerical results corroborating the validity of the DGC criterion. We consider a one-dimensional lattice of \mbox{spin-$s$} sites with the Ising Hamiltonian, with ferromagnetic interaction and transverse field~$h>0$,
\begin{equation}
	 H(s,h) = -\frac{1}{s^2} \sum_{j=1}^L S^{x}_j S^{x}_{j+1} + \frac{h}{s} \sum_{j=1}^L S^{z}_j,
	\label{eq:Isingclass} 
\end{equation}
where $S^{\mu}_j$ are the spin-$s$ matrices at site $j$, satisfying $[S^{k}_{j},S^{l}_{j'}] = i S^{m}_{j} \varepsilon_{klm} \delta_{jj'}$ and ${S^{x}_j}^2 + {S^{y}_j}^2 + {S^{z}_j}^2 = s(s+1) \Id$, with $\hbar=1$ henceforth. The prefactors $1/s$ and $1/s^2$ ensure that the whole class of Hamiltonians $H(s,h)$ yields exactly the same mean-field treatment for all $s$ (see Appendix~\ref{sec:SBMF}). We carry out simulations for the model in Eq.~\eqref{eq:Isingclass} using DMRG for Tree Tensor Networks for ground-state properties~\cite{White1992,MPSAge,Gerster2014b}, and the Time-Evolving Block Decimation (TEBD) algorithm~\cite{Vidal2004,TDMRGWhite} featuring RSVD-compression~\cite{Tamascelli2015RSVD,Kohn2018RSVD} for out-of-equilibrium dynamics, respectively. We adopt a Tensor Network (TN) encoding which protects the $Z_2$ parity symmetry
\begin{equation}
	\Pi = \exp\left( i \pi \sum_ {j=1}^L (s - S^z_j) \right) \; .
\end{equation}
The system size~$L$ in the simulations is chosen large enough to guarantee that finite size effects do not affect the presented results.

\subsection{Equilibrium simulations}

We perform equilibrium simulations to characterize the phase diagram for all $s$, in order to detect the DGC point $h_{\times}(s)$, in addition to the critical point $h_c(s)$. 
While the critical exponents $\nu=1$ and~$z=1$ are independent of $s$ in proximity of~$h_c$, it can be shown that order correlations scale as~$1/s$ (see Appendix~\ref{sec:analytical-hp}). Moreover, $h_c$ increases monotonically with~$s$, with extrema at the limiting cases $h_c(1/2)=1$ and $h_c(\infty)= h^\text{MF}_c = 2$, where $h^\text{MF}_c$ is the critical point of the mean-field treatment of the model, which is independent of $s$ (see Appendix~\ref{sec:SBMF}). The exact form of the dependence of the deviation from the mean-field value $\varepsilon(s) = h^\text{MF}_c - h_c(s)$ on the strength of the quantum fluctuations has been shown to be given by~\cite{Podolsky2014}
\begin{equation}
	\varepsilon(s) =  \frac{\tilde{a}}{s} \left( \left| \ln s \right| - \tilde{b} \right) \; ,
\end{equation}
where $\tilde{a}$, $\tilde{b}$ are non-universal fit constants. In Fig.~\ref{fig:phasediagram} we numerically verify this behavior by plotting the location of the critical points for various values of $s$, together with the fitted function. The resulting fit parameters are $\tilde{a}\approx 0.28$ and $\tilde{b} \approx -2.4$. 
Additionally, we highlight the critical region by plotting the von Neumann entropy $S_\text{VN}(\rho_j)$ of the single-body reduced density matrix~$\rho_j$, in the paramagnetic phase: We observe that only inside the critical region the entropy grows above $10\%$. Finally, Fig.~\ref{fig:phasediagram} contains the location of the DGC points, obtained from the condition $\xi(h_\times) = a = 1$. Here, $\xi$ is the correlation length derived from the ferromagnetic correlation matrix $C_{j,k} = \langle S_j^x S_k^x \rangle/s^2$. We numerically estimate~$\xi$ via
\begin{equation}
	\xi = \sqrt{\sum_{r=1}(r-1)^2 C(r) / \sum_{r=1}C(r)} \; ,
	\label{eq:corrlength}
\end{equation}
where
\begin{equation}
	C(r) = \frac{1}{L-r} \sum_{j=1}^{L-r} C_{j,j+r}
\end{equation}	
is the spatially-averaged correlation function~\cite{WhiteHubbard}.
One can show (see Appendix~\ref{sec:analytical-hp}) that $h_\times(s\rightarrow\infty)=2 \cosh(1)$. For finite~$s$, the trend towards this limit value seems to be well approximated by a power-law decay $h_\times(s) = h_\times(s\rightarrow\infty) - \tilde{c} \,s^{- \tilde{\eta}}$, yielding fitted constants $\tilde{c}\approx 0.31$ and $\tilde{\eta}\approx 0.52$.
Remarkably, the DGC delivers a finite interval $[2,2\cosh(1)]$ of the quantum critical region in the quasiclassical limit $s \to \infty$, in contrast to the traditional Ginzburg criterion.

\subsection{Out-of-equilibrium simulations}

\begin{figure}
	\includegraphics[width=\columnwidth]{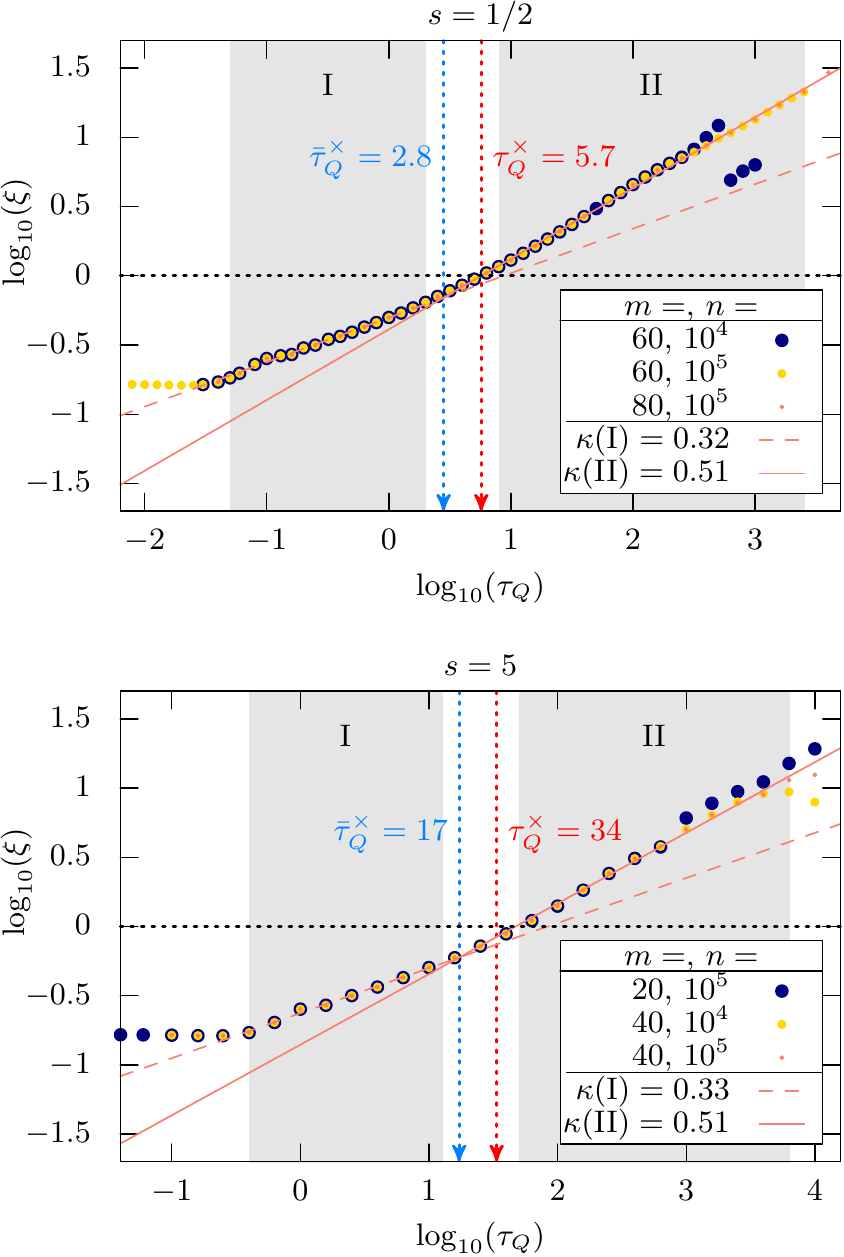}
	\caption{\label{fig:KZcompare}Comparison between estimated and observed KZ crossover quench times, for $L=128$ and two different spins: $s=1/2$ (top panel) and $s = 5$ (bottom panel). Both panels show the crossover between the classical KZ scaling $\kappa \simeq 1/3$ and the quantum KZ scaling $\kappa \simeq 1/2$, which allows us to identify the observed crossover quench time $\bar{\tau}_Q^{\times}$, where the two power-laws intersect. The red arrow shows the estimated crossover quench time, obtained from Eq.~\eqref{eq:taucross} via the DGC. Data points are shown for different bond dimensions~$m$ and numbers of TEBD time steps $n$, in order to demonstrate numerical convergence of the simulation data.}
\end{figure}

We performed numerical simulations of the many-body dynamics generated by the linearly quenched Ising Hamiltonian of Eq.~\eqref{eq:Isingclass}. We considered various values of $s$ and system sizes $L$ of the order of $10^2$ sites, using a fixed quench interval from $h_\mathrm{ini}=30$ (deep in the paramagnetic phase) to $h_\mathrm{fin}=0.5$ (in the ferromagnetic phase). We use the correlation length~$\xi$ of the final state as inverse defect measure.
The results of the simulations, for two different values of the spin quantum number ($s=1/2$ and $s=5$), are reported in Fig.~\ref{fig:KZcompare}. Both scenarios deliver the predicted behavior: For small quench durations, the fitted KZ exponent is very close to~$\kappa = 1/3$, while for long quenches it is very close to~$\kappa=1/2$. The observed crossover quench time~$\bar{\tau}^{\times}_{Q}$ between the two regimes is well approximated by the~$\tau^{\times}_{Q}$ estimated from Eq.~\eqref{eq:taucross}, where $\varphi$ has been determined for the accessible gap (see Appendix~\ref{sec:energy-gap})~\cite{IsingDQPT}.

\begin{figure}
	\includegraphics[width=\columnwidth]{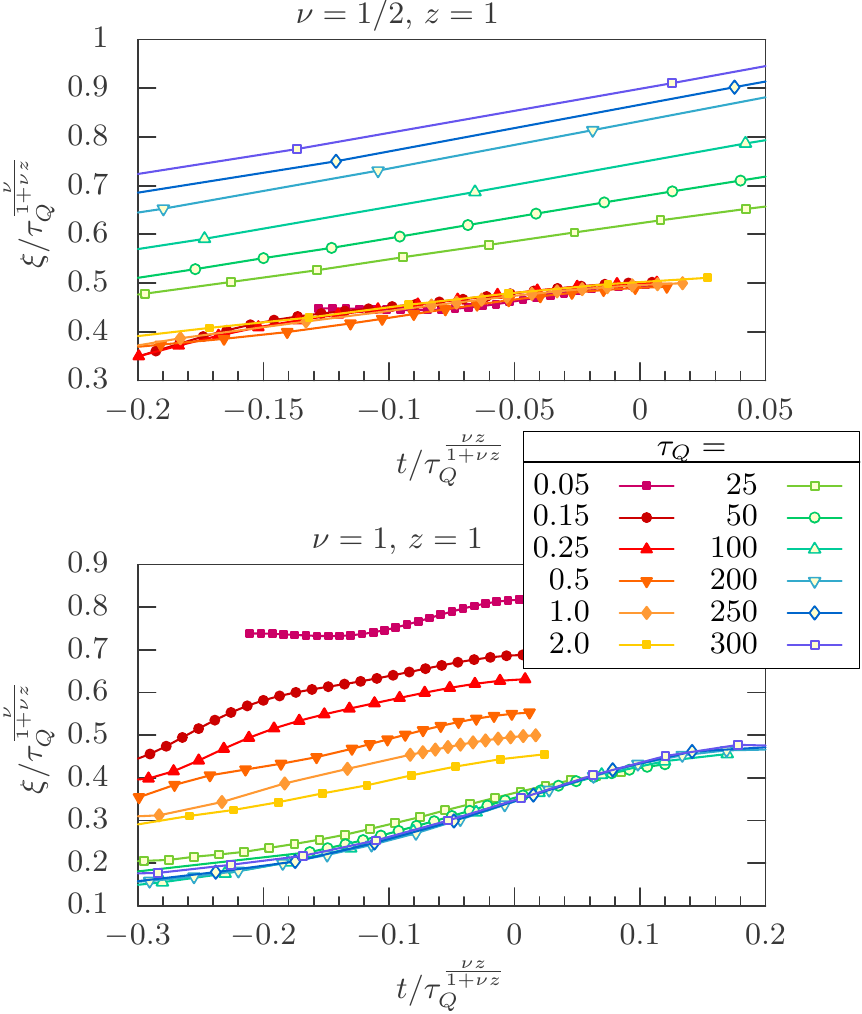}
	\caption{ \label{fig:FTS}Finite-time scaling of the (inverse) density of defects during the quench, for $s=1/2$. The curves show the correlation length as a function of time, for various total quench times~$\tau_Q$, where the axes have been rescaled according to Eq.~\eqref{eq:FTS} for two different sets of critical exponents: the classical ones (top panel) and the quantum ones (bottom panel). As expected, only the curves corresponding to quench times shorter (longer) than the crossover quench time, signaled by warm colors (cold colors), collapse into a main trend, while the other curves are outliers.
	}
\end{figure}

To further strengthen our results, we perform a Finite-Time Scaling (FTS) analysis~\cite{Huang-finite-time-scaling:2014,Polkovnikov-finite-time-scaling:2014}, the out-of-equilibrium analog of the finite-size scaling analysis~\cite{FisherBarberFSS}. Within this framework, we fully embrace the KZ approximation, according to which the evolution is adiabatic until freeze-out, while the order properties stay constant afterwards. In this picture, the time-dependent correlation length $\xi(t)$ during the quench must undergo the following scaling:
\begin{equation}
	\xi(t) \simeq \tau_Q^{\frac{\nu}{1+ \nu z}} f \left( t \; \tau_Q^{-\frac{\nu z}{1 + \nu z}} \right) \; ,
	\label{eq:FTS}
\end{equation}
as long as $\xi < L$, where $f(\cdot)$ is a non-universal function.
This expression guarantees that $\hat{t}$, $\xi(0)$ and $\xi(\hat{t})$ scale with $\tau_Q$ with the correct KZ exponents.
In Fig.~\ref{fig:FTS} we observe a collapse of the curves $\xi(t)$ according to Eq.~\eqref{eq:FTS}. Again, we observe excellent agreement with our predictions: When using the quantum critical exponents $z = \nu = 1$ (classical critical exponents $z = 2 \nu =1$) we observe a collapse only of the curves with quench times longer (shorter) than the estimated crossover $\tau_Q > \tau_Q^{\times}$ ($\tau_Q <\tau_Q^{\times}$), while the other curves being clear outliers.

\section{Conclusion}

We proposed a general, yet simple criterion based on quantitative equilibrium properties to predict the timescale at which the crossover between a classical KZ scaling of defects, and a quantum KZ scaling, is expected to occur for linear quenches on nearest-neighbor interacting lattice models. Our DGC simply discriminates whether the correlation length at freeze-out is longer or shorter than the lattice constant, resulting in a quantum or classical scaling respectively. We tested our conjecture on the spin-$s$ quantum Ising model class in 1D, and observed remarkable agreement with the DGC estimation.

This study puts more solid ground on the phenomenon of the quantum--classical KZ crossover. Moreover, the DGC criterion is a ready-to-use estimator, for any quantum lattice experiment of quench dynamics, to quickly understand whether the quantum KZ regime is accessible within its experimental specifications. We estimate that our conjecture could be readily verified experimentally on atomic quantum-simulator platforms, such as analog quantum simulators on trapped ion architectures~\cite{PorrasCirac_AQS,RoosBlatt_AQS}, or Rydberg atoms trapped in arrays of optical tweezers~\cite{51Rydbergs} where recently the first observation of a genuinely quantum KZ mechanism was made~\cite{51RydbergsKZ}.

\begin{acknowledgments}

We thank L. Mazza, D. Podolsky, R. Fazio and G. Morigi for feedback and discussions. 
Numerical calculations have been performed with the computational resources provided by the bwUniCluster project~\footnote{{b}wUniCluster: funded by the Ministry of Science, Research and Arts and the universities of the state of Baden-W{\"u}rttemberg,  Germany, within the framework program bwHPC.}. We acknowledge support from the Baden-W\"urttemberg Stiftung via Eliteprogramm for PostDocs, the DFG via Heisenberg fellowship and TWITTER project, the EU via UQUAM, QUANTERA via \mbox{QTFLAG},
the US Air Force Office of Scientific Research (AFOSR) grant FA9550-18-1-0319,
the Austrian Research Promotion Agency (FFG) contract 872766 AutomatiQ,
and
the Carl-Zeiss-Stiftung via Nachwuchsf\"orderprogramm.

\end{acknowledgments}

\appendix

\section{Mean-field equivalence of the spin-$s$ Ising model} \label{sec:SBMF}

In this section, we show that the Single-Body Mean-Field (SBMF) solution of the spin-$s$ Ising model in 1D, given by Eq.~\eqref{eq:Isingclass}, is independent of $s$. In the SBMF ansatz, the reduced density matrices decompose into their single-body components $\rho_{j,j'} = \rho_j \otimes \rho_{j'}$, which will be homogeneous $\rho_j = \rho_{j'} \,  \forall j,j'$ since the ferromagnetic interaction we consider does not spontaneously break translational invariance. We now prove that, regardless of the spin $s$, the critical point is always at $h^{\text{MF}}_{c} = 2$ and the order parameter is $ \langle O \rangle = \frac{1}{2} \sqrt{4-h^2}$ for $h \leq 2$ while $\langle O \rangle = 0$ for $h \geq 2$, where $O = \frac{1}{s} S^x$. To do this, we first derive the spin-$1/2$ solution and then show that larger spins lead to an analogous classical minimization functional.

\subsection{Spin-1/2}

In this scenario, we explicitly write the single-spin density matrix $\rho_j = \frac{1}{2} \Id + \frac{\vec{r}}{2} \cdot \vec{\sigma}$, where now $\vec{\sigma} = \frac{1}{s} \vec{S}$ is the vector of Pauli matrices, and $O = \sigma^x$. Positivity of the density matrix requires $| \vec{r} | \leq 1$, and clearly $\langle \sigma^x \rangle = r_x$ while $ \langle {\sigma}^{z} \rangle = r_z$. In order to respect the bound $r_x^2 + r_z^2 \leq |\vec{r}| \leq 1$ we use the parametrization $r_x = r \cos {\theta}$ and $r_z = r \sin{\theta}$, with $r \in [0,1]$. The SBMF functional to minimize then reads
\begin{multline} \label{eq:functional} 
	\langle H(1/2,h)\rangle_{\text{MF}} = 
	- \frac{1}{s^2} \langle S^x \rangle^2 + \frac{h}{s} \langle S^{z} \rangle = \\
	- \langle \sigma^x \rangle^2 + h \langle \sigma^{z} \rangle
	= -  r^2 \cos^2{\theta} + hr \sin{\theta} \; ,
\end{multline}
The solution will definitely be in the interval $\theta \in [-\pi, 0]$, since for any $\theta$ value within $[0, \pi]$, the angle $\theta' = -\theta$ returns an equal or better value of the functional. Within this interval, both summands in the expression~\eqref{eq:functional} will be negative. Therefore, the global minimum will be at $r = 1$, and the coordinates of the optimal solution can be given analytically:
\begin{eqnarray} \label{eq:mfsol}
	r_\mathrm{min} &=& 1 \; , \\
	\theta_\mathrm{min} &=& 
	\begin{cases}
		 - \arcsin\left(\frac{h}{2}\right),  \quad &\text{for} \quad 0 \leq h \leq 2 \\
		 - \frac{\pi}{2} \, ,  \quad &\text{for} \quad h \geq 2 \; ,
	\end{cases} \notag
\end{eqnarray}
while the minimized energy functional is equal to
\begin{equation} 
  \langle H(1/2,h)\rangle_{\text{MF}}^{\mathrm{min}} =
  \begin{cases}
   -1 - \frac{h^2}{4} \, , \quad &\text{for} \quad 0 \leq h \leq 2\\
   -h \, , \quad &\text{for} \quad h \geq 2,
 \end{cases}
\end{equation}
and the order parameter is $ \langle O \rangle = \frac{1}{2} \sqrt{4-h^2}$.
Interestingly, the corresponding critical exponent~$\beta$, which relates to the
spontaneous local order $\langle O \rangle \sim (h_c - h)^{\beta}$, corresponds to $\beta = 1/2$ for the SBMF transition, in contrast to the known $\beta = 1/8$ of the full quantum treatment~\cite{Onsager44IsingModel}.

\subsection{Spin-$s$}
Here we show that the SBMF treatment leads to minimizing a functional equivalent to Eq.~\eqref{eq:functional}.
We first prove that $|\langle \vec{S} \rangle |^2 \leq s^2$, which is strictly smaller than
$\langle | \vec{S} |^2 \rangle =  s(s+1)$.  This is seen by setting $\vec{a} = \langle \vec{S} \rangle$ and then noticing that
\begin{equation}
	\vec{a} \cdot \vec{a} = \left( \vec{a} \cdot \frac{\vec{a}}{|\vec{a}|} \right)^2 =  \langle \vec{S} \cdot \vec{a}/|\vec{a}| \rangle^2 \; .
\end{equation}
Since now $\vec{a}/|\vec{a}|$ is a vector of modulus one, we know that $\vec{S} \cdot \vec{a}/|\vec{a}|$ is a rotated spin-$s$ matrix, and its spectrum is between $-s$ and $s$. It follows that $-s \leq \langle \vec{S} \cdot \vec{a}/|\vec{a}| \rangle \leq s$, and therefore $\langle \vec{S} \cdot \vec{a}/|\vec{a}| \rangle^2 = | \langle \vec{S} \rangle|^2 \leq s^2$. This means that
\begin{equation}
	\frac{1}{s^2} \langle S^x \rangle^2 + \frac{1}{s^2} \langle S^z \rangle^2 \leq 1
\end{equation}
regardless of~$s$. And since Eq.~\eqref{eq:mfsol} is the most generic solution of the functional~\eqref{eq:functional} under this constraint, we can conclude that spin-$s$ cannot exhibit a better solution than Eq.~\eqref{eq:mfsol}.
Moreover, let us now show that this solution exists for every~$s$:
Specifically, we consider the spin-$s$ subclass of states
\begin{equation}
	|\theta \rangle = e^{i (\pi/2 - \theta) S^{y}} |m=+s\rangle
\end{equation}
parametrized by $\theta \in [0,2\pi]$. These states exhibit by construction $\langle {S}^{x} \rangle = s \cos \theta$ and $\langle {S}^{z} \rangle = s \sin \theta$. The solution given by Eq.~\eqref{eq:mfsol} thus exists and minimizes the SBMF functional, which makes it the minimal solution for all~$s$.

\section{Procedure for determining the critical point} \label{sec:criticalpoints}

\begin{figure}
	\includegraphics[width=0.97\columnwidth]{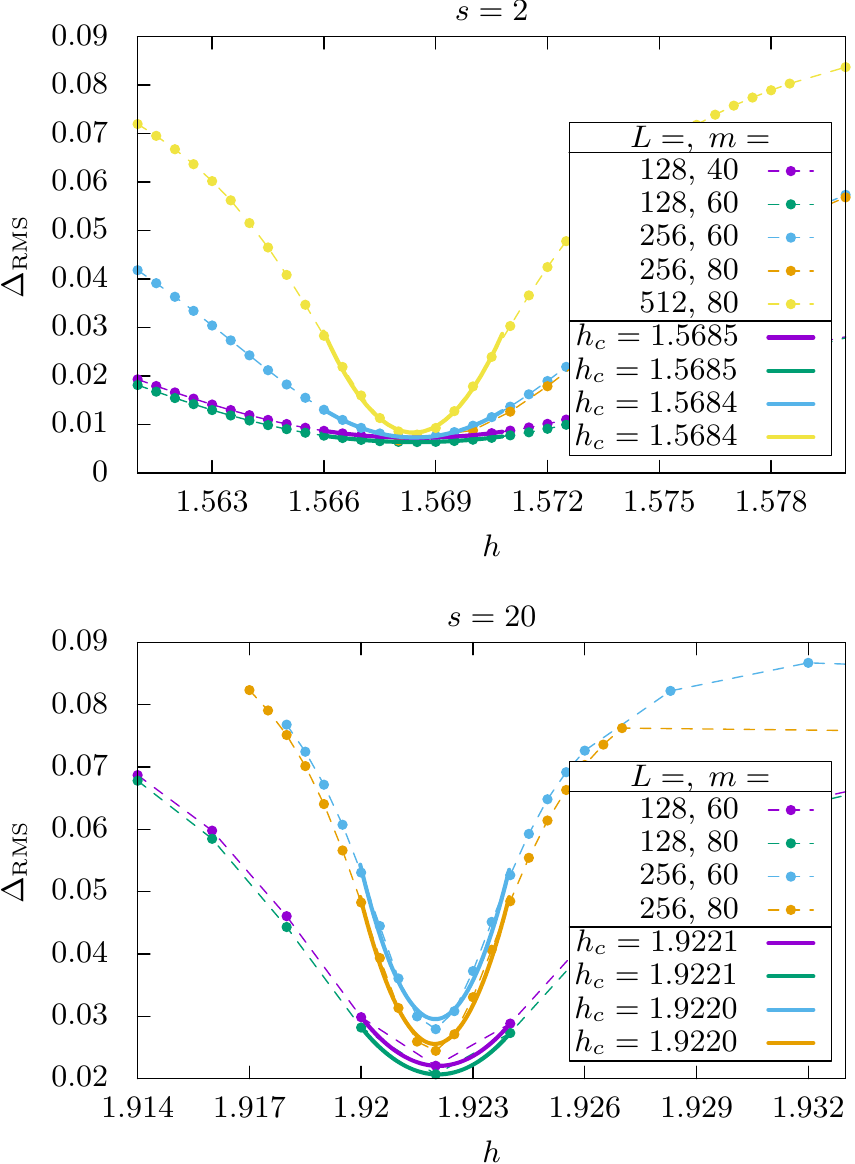}
	\caption{\label{fig:hcrit-determination}
		Determination of the critical field strength~$h_c$ for $s=2$ (top) and $s=20$ (bottom), using different system sizes~$L$ and TN bond dimensions~$m$.}
\end{figure}

The critical points $h_c$ shown in Fig.~\ref{fig:phasediagram} haven been numerically obtained from TN simulations via the following procedure: Precisely at the critical point, the von Neumann entropy in a system of size~$L$ with periodic boundary conditions (PBC) is known to scale like~\cite{Calabrese2004}
\begin{equation}
	S_\text{VN}(\ell) = \frac{c}{3} \log_2 \left[ \mathrm{crd}(\ell) \right] + c^\prime_1  \; , 
	\label{eq:critentropy}
\end{equation}
as a function of the partition size $\ell$, with $\mathrm{crd}(\ell)=L/\pi \, \sin(\pi\ell/L)$. Here, $c$ is the conformal central charge (for the Ising universality class we have $c=1/2$), and $c^\prime_1$ is a non-universal constant. The strategy is now to fit the numerical data to this expression for various values of the field strength $h$, in order to probe agreement with the critical scaling. The value of~$h$ where the fit displays maximal agreement, quantified by the fit's root mean square deviation $\Delta_\mathrm{RMS}$, represents the location of the critical point $h_c$. This procedure is shown in Fig.~\ref{fig:hcrit-determination}, for two different values of~$s$, and various system sizes~$L$ and TN bond dimensions~$m$.

\section{Analytical solution for large $s$ via Holstein--Primakoff transformation} \label{sec:analytical-hp}

\begin{figure}
	\includegraphics[width=0.95\columnwidth]{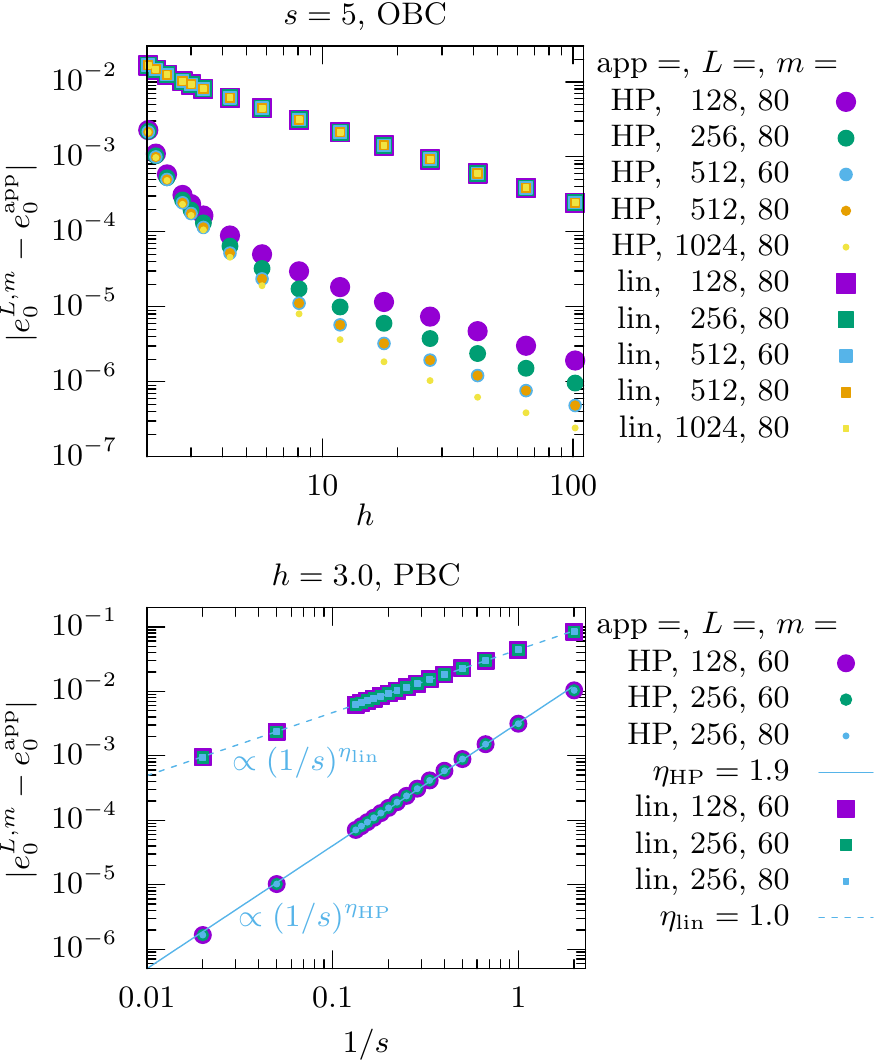}
	\caption{\label{fig:gsenergy-HP}Comparison of the numerically determined GS energies in the paramagnetic phase with the large-$s$ HP result Eq.~\eqref{eq:gsenergy_hp}, and with the GS energies from the linear approximation $e_0^\text{lin}=-h$. The TN simulations have system size~$L$, bond dimension~$m$, and either open (OBC) or periodic boundary conditions (PBC).}
\end{figure}
\begin{figure}
	\includegraphics[width=\columnwidth]{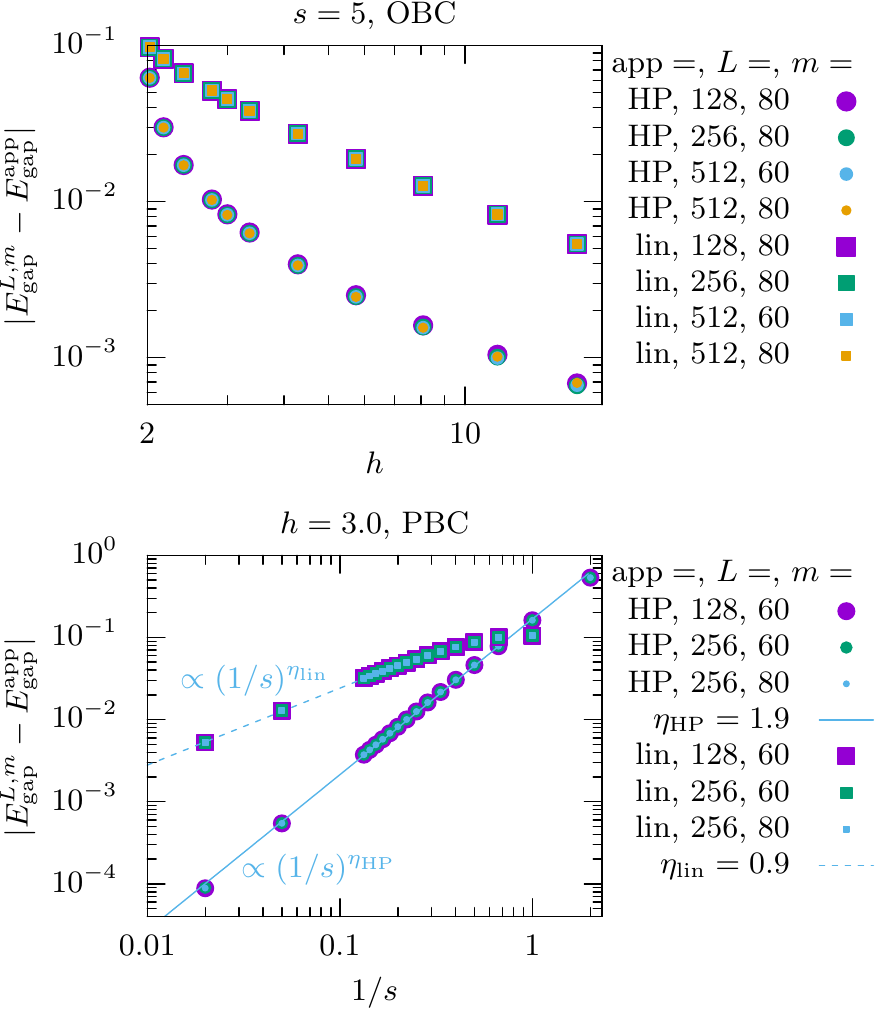}
	\caption{\label{fig:energygap-HP}Comparison of the numerically determined energy gaps with the large-$s$ HP result Eq.~\eqref{eq:energygap_hp}, and with the gaps predicted by the linear approximation $E_\text{gap}^\text{lin}=(h-1)/s$.}
\end{figure}

We employ the Holstein--Primakoff (HP) transformation~\cite{Holstein1940}
\begin{eqnarray}
	S_j^z &=& a^\dagger_j a^{}_j - s  \notag \\
	S_j^+ &=& \sqrt{2s} \, a^\dagger_j \sqrt{1-a_j^\dagger a_j/ (2s)} \notag \\
	S_j^- &=& \sqrt{2s} \, \sqrt{1-a_j^\dagger a_j/ (2s)} \; a_j \; ,
	\label{eq:hp-trans}
\end{eqnarray}
with $S^{\pm}_j = S^x_j \pm i S^y_j$ the raising and lowering operators as usual, and $a^{}_j$ ($a^\dagger_j$) is a bosonic annihilation (creation) operator. We expand the square roots in Eq.~\eqref{eq:hp-trans} to lowest order:
\begin{equation}
	S_j^+ \simeq \sqrt{2s} \; a_j^\dagger \; , \qquad S_j^- \simeq \sqrt{2s} \; a_j \; , 
	\label{eq:hptrans-lowestorder}
\end{equation}
which is a good approximation for sufficiently large $s$. Note that this transformation from a finite- to an infinite-dimensional Hilbert space is only faithful for states which populate exclusively one end of the level spectrum of~$S_j^z$. Thus, for the spin-$s$ Ising model, this transformation is only useful in the paramagnetic phase, while in the ferromagnetic phase it fails to preserve the physics of the model. Via this transformation, we obtain from the original spin Hamiltonian defined in Eq.~\eqref{eq:Isingclass} the following bosonic quadratic Hamiltonian:
\begin{eqnarray}
	H_\mathrm{HP} &=& -\frac{1}{2s} \sum_{j=1}^L \left( a_j a_{j+1}  + a^{}_j a^\dagger_{j+1} \right)  + \mathrm{h.c.} \notag \\  
		&+& \frac{h}{s} \sum_{j=1}^L a_j^\dagger a_j - L \, h \; .
	\label{eq:hamiltonian-hp}
\end{eqnarray}
In order to diagonalize $H_\mathrm{HP}$, we first perform a transformation to $k$-space, using a new set of bosonic operators:
\begin{eqnarray}
	\tilde{a}_k &=& \frac{1}{\sqrt{L}} \sum_{j=1}^L e^{-i k j} \, a_j \:, k \in \left\{\frac{2\pi}{L} m \mid m=-\frac{L}{2} ... \frac{L}{2}-1  \right\} \notag \\
	a_j &=& \frac{1}{\sqrt{L}} \sum_k e^{ikj} \, \tilde{a}_k
	\label{eq:fourier-trans}
\end{eqnarray}
After applying this transformation, the Hamiltonian Eq.~\eqref{eq:hamiltonian-hp} becomes:
\begin{eqnarray}
	\tilde{H}_\mathrm{HP} &=& \frac{1}{s} \sum_k \left( h - \cos(k) \right) \tilde{a}^\dagger_k \tilde{a}^{}_k \notag \\
		&-& \frac{1}{2s} \sum_k \left( e^{-ik} \, \tilde{a}_k \tilde{a}_{-k} + \mathrm{h.c.} \right) - L \, h
	\label{eq:hamiltonian-hp-kspace}
\end{eqnarray}
Finally, we use a Bogoliubov transformation~\cite{Bogoliubov1947}
\begin{eqnarray}
	b_k &=& \cosh(\phi) \, \tilde{a}^{}_k - \sinh(\phi) \, \tilde{a}^\dagger_{-k} \notag \\
	\tilde{a}_k &=& \cosh(\phi) \, b^{}_k + \sinh(\phi) \, b^\dagger_{-k} \; ,
	\label{eq:bog-trans}
\end{eqnarray}
which diagonalizes $\tilde{H}_\mathrm{HP}$, if the parameter $\phi$ is chosen such that it satisfies the relation $\tanh(2\phi) = \cos(k)/[h-\cos(k)]$. The resulting diagonal Hamiltonian then reads
\begin{eqnarray}
	H_\mathrm{Bog} &=& \frac{h}{s} \sum_k \sqrt{1 - 2\cos(k)/h} \; b^\dagger_k b^{}_k \nonumber\\
		&-& \frac{h}{4s} \sum_k \left(  \sqrt{1-2\cos(k)/h} - 1 \right)^2 \nonumber\\
		&-& L \, h \; .
	\label{eq:hamiltonian-hp-bog}
\end{eqnarray}
From $H_\mathrm{Bog}$ one immediately obtains the ground state (GS) energy per site $e_0 = E_0/L$. For $L \rightarrow \infty$, {\it i.e.\@} in the thermodynamic limit, it reads
\begin{equation}
	e_0 = \frac{1}{s} \left( \frac{1}{\pi} \sqrt{h(h+2)} \; E[4/(h+2)] - \frac{h}{2} \right) - h \; ,
	\label{eq:gsenergy_hp}
\end{equation}
where $E[x]$ is the complete elliptic integral of the second kind
\begin{equation}
	E[x]=\int_0^{\pi/2} \sqrt{1-x \sin^2(\theta)} \, \mathrm{d}\theta \; .
\end{equation}
Fig.~\ref{fig:gsenergy-HP} shows a comparison of the expression Eq.~\eqref{eq:gsenergy_hp} with numerically determined GS energies in the paramagnetic phase. We observe (see Fig.~\ref{fig:gsenergy-HP}) that the GS energies obtained from the HP approximation have an error of order $\mathcal{O}(1/s^2)$, {\it i.e.\@} they are the next-order correction to the linear GS energy $e_0^\text{lin}=-h$, which is exact in the limit $h\rightarrow\infty$ and has an error of $\mathcal{O}(1/s)$ for finite $h$.

\begin{figure}
	\includegraphics[width=0.95\columnwidth]{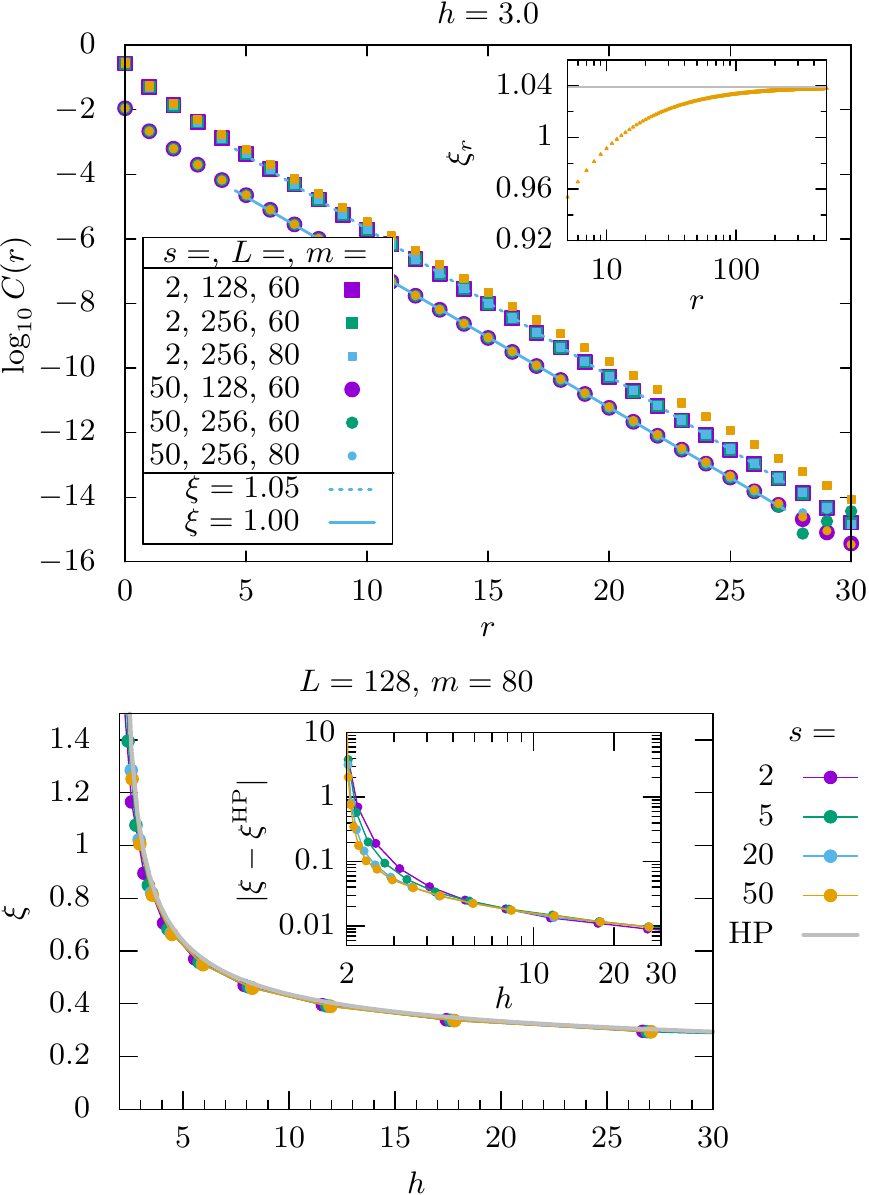}
	\caption{\label{fig:corrlength-HP}Top panel: Comparison of correlation functions from TN simulations with the large-$s$ HP result Eq.~\eqref{eq:corrfunc_hp} (orange points), for fixed field strength $h=3$ and two different spin quantum numbers~$s=2$ (squares) and $s=50$ (circles). The inset shows how $\xi_r$ (as defined in Eq.~\eqref{eq:corrlen_corrfunc-ratio}) approaches the constant $[\mathrm{arcosh}(h/2)]^{-1}$ (gray line) for $r\rightarrow\infty$. Bottom panel: Comparison of the HP correlation length Eq.~\eqref{eq:corrlength_hp} with TN correlation lengths, as a function of the field strength $h$ and for various $s$. The inset shows the same data, but now plotting the differences between the TN curves and the HP curve.}
\end{figure}

According to $H_\mathrm{Bog}$, the energy gap to the first excited state is
\begin{equation}
	E_\text{gap} = \frac{h}{s} \, \sqrt{1-2/h} \; .
	\label{eq:energygap_hp}
\end{equation}
Fig.~\ref{fig:energygap-HP} shows a comparison of this expression with numerical data, demonstrating again its improved accuracy over the expression $E_\text{gap}^\text{lin}=(h-1)/s$, valid in the limit $h \rightarrow\infty$.

The sequence of transformations outlined above also allows to calculate the GS correlation function~$C(r)$. This can be achieved by considering the expectation value $\langle \Psi_0 | \sum_j S^x_j S^x_{j+r}  | \Psi_0 \rangle/L s^2$, where $|\Psi_0\rangle$ is the GS. After again transforming the spin operators to the set of Bogoliubov operators $\{b^{}_k\}$, $\{b^\dagger_k \}$, one readily obtains (for $L\rightarrow\infty$):
\begin{equation}
	C(r) = \frac{1}{2s} \frac{1}{2\pi} \int_{-\pi}^\pi \frac{\cos(rk)}{\sqrt{1-2\cos(k)/h}} \, \mathrm{d}k \; .
	\label{eq:corrfunc_integral_hp}
\end{equation}
The solution of this integral can be written as the following series:
\begin{eqnarray}
	\label{eq:corrfunc_hp}
	C(r) &=& \frac{1}{2s} \sqrt{\frac{h}{h+2}} \\ 
		&\times& \sum_{n=r}^\infty \frac{ \left[ (2n)! \right]^2 }{(n-r)! \, (n+r)! \, \left[ n! \right]^2} \left( \frac{1}{4(h+2)} \right)^n \nonumber
\end{eqnarray}
A comparison of this expression with correlation functions obtained from TN simulations is shown in the top panel of Fig.~\ref{fig:corrlength-HP}, for a fixed field strength $h=3$. As expected, the larger the spin quantum number $s$, the better the agreement. Or, in other words: Eq.~\eqref{eq:corrfunc_hp} becomes exact (but also trivial) for $s\rightarrow\infty$. One can show that the ratio $C(r)/C(r+1)$ approaches a constant for $r\rightarrow\infty$, allowing one to calculate the correlation length 
\begin{equation}
	\xi_r = \frac{1}{\log \left[C(r)/C(r+1)\right]}
	\label{eq:corrlen_corrfunc-ratio}
\end{equation}
by taking the limit $r\rightarrow \infty$. This leads to
\begin{equation}
	\xi(h) = \frac{1}{\log\left[ h/2 + \sqrt{h^2/4 - 1} \right]} = \frac{1}{\mathrm{arcosh}(h/2)} \; .
	\label{eq:corrlength_hp}
\end{equation}
Note that this expression for $\xi(h)$ does not contain~$s$, meaning that our lowest-order expansion of the HP transformation fails to capture the $s$-dependence of the correlation length. Nevertheless, Eq.~\eqref{eq:corrlength_hp} is still quite useful because the convergence of the numerical correlation lengths (determined from TN simulations) to Eq.~\eqref{eq:corrlength_hp} is rather fast with increasing~$s$: This is demonstrated in the bottom panel of Fig.~\ref{fig:corrlength-HP}, where TN data is compared to Eq.~\eqref{eq:corrlength_hp}. Moreover, Eq.~\eqref{eq:corrlength_hp} allows us to easily calculate the DGC point $h_\times$ in the limit $s\rightarrow\infty$: Since by definition $\xi(h_\times)=1$, we immediately arrive at $h_\times(s\rightarrow\infty)=2\cosh(1)$.

\section{Equilibrium properties of the spin-$s$ Ising Model} \label{sec:equilibrium-properties-spinS-Ising}

\begin{figure}
	\includegraphics[width=\columnwidth]{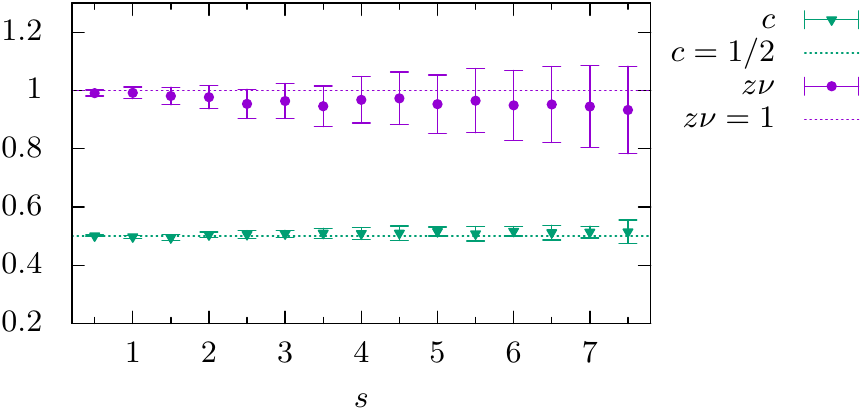}
	\caption{\label{fig:universality}Universality of the phase transition: numerically determined central charge $c$ and energy gap exponent $z\nu$ as a function of the spin quantum number $s$.}
\end{figure}

Here we discuss in more detail the (zero temperature) equilibrium properties of the spin-$s$ Ising model as defined in Eq.~\eqref{eq:Isingclass}. Because of
\begin{equation}
\left[ \frac{S_j^a}{s}, \frac{S_{j'}^b}{s} \right] = i \delta_{jj'} \varepsilon_{abc} \, \frac{S_j^c}{s} \, \frac{1}{s} \; , \quad a,b,c \in \{x,y,z\} \; ,
\label{eq:commutator}
\end{equation}
the quantum frustration of the Hamiltonian terms decreases for increasing~$s$, and the inverse spin $1/s$ can be interpreted as an ``effective~$\hbar$". For $s\to\infty$ this effective~$\hbar$ vanishes. The ``rescaled" spin operators $S^a/s$ have a bounded spectrum of equispaced eigenvalues in the interval $[-1,1]$, which in the limit $s\rightarrow\infty$ becomes continuous. These observations justify the statement that for $s\to\infty$ the spin-$s$ Ising model becomes quasiclassical: All operators commute with each other, and the quantization of expectation values disappears. Via mean-field theory, which becomes exact for infinitely large $s$, it can be shown (see Appendix~\ref{sec:SBMF}) that for $s\to\infty$ the model has a critical point at $|h_c|=2$, separating the ferromagnetic phase $|h|<2$ with non-vanishing ferromagnetic local order parameter 
\begin{equation}
	M = \sqrt{\frac{1}{L(L-1)} \; \sum_{j\neq k} \frac{\langle S^x_j S^x_k \rangle}{s^2}}
	\label{eq:magnetization}
\end{equation}
from the paramagnetic phase $|h| > 2$ with vanishing $M$ at $L \to \infty$. The other limiting case, namely $s=1/2$, can also be solved analytically, via a mapping to free fermions~\cite{Lieb1961}. The quantum phase transition in this case occurs at $|h_c|=1$. For all other finite values of~$s$ we resort to numerical Tensor Network (TN) simulations based on the DMRG algorithm~\cite{MPSAge,TNAnthology} in order to determine the critical point and other quantities of interest.

\subsection{Quantum phase transition and critical behavior}

\begin{figure}
	\includegraphics[width=\columnwidth]{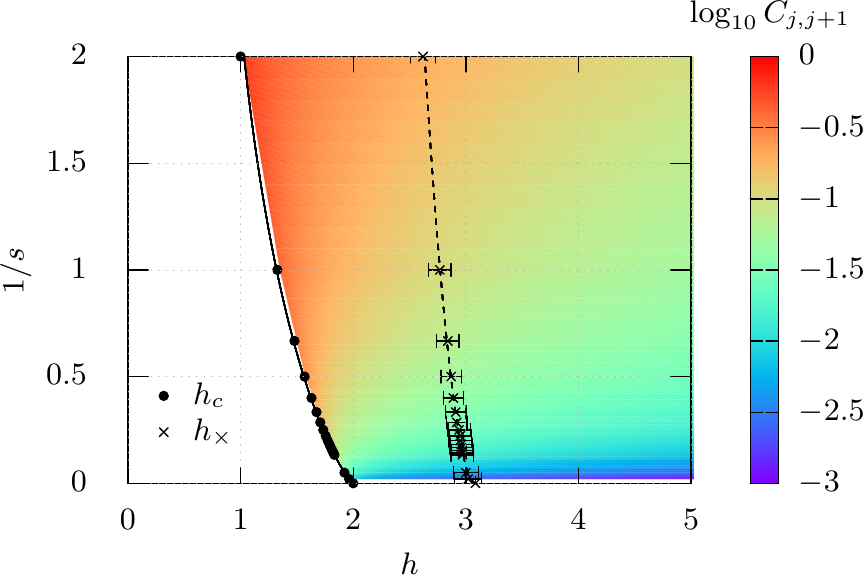}
	\caption{\label{fig:phasediag_corr}Phase diagram as a function of the transverse field~$h$ and the inverse spin $1/s$: The critical points $h_c$ separate the ferromagnetic phase ($h<h_c$) from the paramagnetic phase ($h>h_c$). The color in the paramagnetic phase indicates the magnitude of the nearest-neighbor correlations, visualizing the vanishing of quantum fluctuations for $s\to\infty$ or $h\to\infty$. The crosses mark the values $h_\times$ at which the correlation length $\xi$ equals one lattice site, {\it i.e.\@} $\xi(h \lessgtr h_\times) \gtrless 1$.}
\end{figure}

We start by characterizing the quantum phase transition of the model, occurring for all finite values of $s$. As a consequence of scale invariance in the vicinity of the quantum critical point, the physics of the model at the transition is insensitive to microscopical details. It is therefore completely determined by its underlying conformal field theory, which in turn is determined by the model's universality class. The universality class of a model only depends on the symmetries that are broken at the phase transition, and the dimensionality of the model. Since the broken symmetry of the Ising model is always $Z_2$, and we are always working in one spatial dimension, it is to be expected that the critical properties of the model do not depend on $s$. In particular, the critical exponents $\nu$ and~$z$ (determining the power-law scalings of the correlation length $\xi \propto |h-h_c|^{-\nu}$ and of the energy gap $E_\text{gap} \propto |h-h_c|^{z\nu}$), as well as the central charge $c$, should be constant. In Fig.~\ref{fig:universality} we verify that this is indeed the case: For all values of $s$, the numerically determined values of the aforementioned quantities are compatible with $\nu=1$, $z=1$, and $c=1/2$, corresponding to the so-called Ising universality class.

On the other hand, we have argued above that the strength of the quantum fluctuations (the ``effective $\hbar$") is proportional to $1/s$. This means that the interval around the critical point where quantum fluctuations are predominant (critical region) is shrinking for increasing~$s$. Another immediate consequence of reduced quantum fluctuations is a shift of the critical point $h_c$ towards larger values: The smaller the quantum fluctuations, the larger the transverse field strength $h_c$ required to completely destroy the ferromagnetic order. In Fig.~\ref{fig:phasediag_corr}, we show the shrinking of the critical region on the paramagnetic side of the phase diagram by plotting the nearest-neighbor correlations $C_{j,j+1}=\langle S^x_j S^x_{j+1} \rangle/s^2$. This serves as a witness of quantum fluctuations because only their presence allows $C_{j,j+1}$ to be non-vanishing in the paramagnetic phase.
An alternative way to evidence quantum fluctuations, namely via an entanglement measure, is given by the color plot in Fig.~\ref{fig:phasediagram}. There, the von Neumann entropy $S_\text{VN}(\rho_j)$ of the single-body density matrix $\rho_j$  is plotted. More precisely, 
\begin{equation}
\rho_j = \mathrm{Tr}_{\{1,\ldots,L\}\setminus j} |\Psi_0\rangle \langle \Psi_0| \; ,
\end{equation}
where $|\Psi_0\rangle$ is the ground state of the spin-$s$ Ising Hamiltonian, and the trace runs over all sites except $j$.

\subsection{Behavior of the correlation length}

\begin{figure}
	\includegraphics[width=0.95\columnwidth]{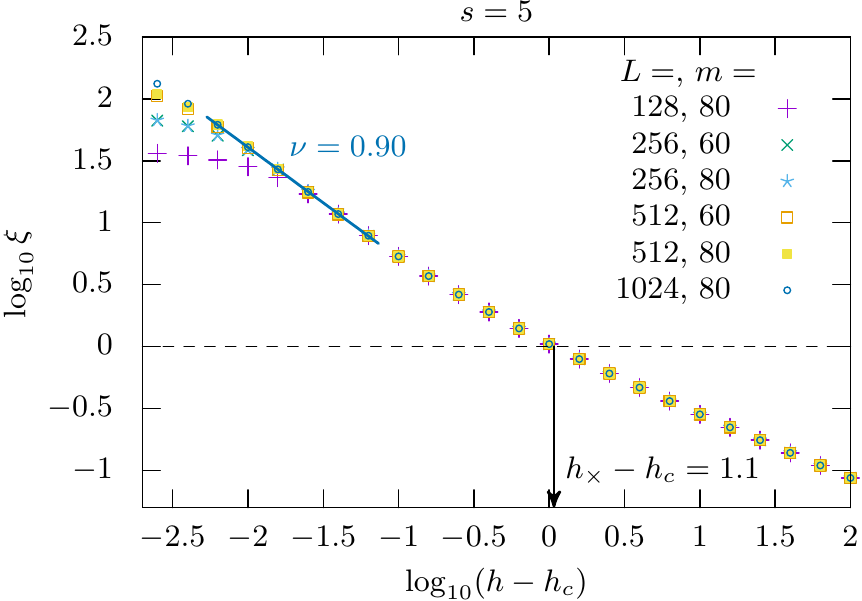}
	\caption{\label{fig:corrlength}Numerically determined correlation lengths in the paramagnetic phase, for $s=5$ and various system sizes $L$ and bond dimensions $m$. The blue line is a power-law fit in order to determine the critical exponent $\nu$ close to the phase transition.}
\end{figure}

Valuable information on the spatial extent of correlations of a given ground state is provided by its correlation length $\xi$. It can be obtained by considering the two-site correlations  $C_{j,k} = \langle S_j^x S_k^x \rangle/s^2$ and the corresponding correlation function $C(r) = C_{j,j+r}$. In the paramagnetic phase this correlation function decays exponentially, {\it i.e.\@} according to $C(r) \propto \exp(-r/\xi)$, for $r$ large enough. In Fig.~\ref{fig:corrlength} we show numerically determined correlation lengths $\xi(h)$, using the example $s=5$.
Close to the phase transition, {\it i.e.\@} for $|h-h_c| \ll 1$, this data can be used to determine the critical exponent $\nu$. The numerically determined value $\nu \approx 0.9$ is indeed compatible with the quantum prediction $\nu = 1$. On the other hand, far from the phase transition, $\xi$ tends to zero. Based on the definition of the DGC outlined in Sec.~\ref{sec:DGC}, we determine $h_\times$ via the condition $\xi(h_\times)=1$: For $h>h_\times$, quantum correlations are negligible and the ground state of the model is very similar to a classical paramagnet.

\subsection{Behavior of the energy gap} \label{sec:energy-gap}

\begin{figure}
	\includegraphics[width=0.94\columnwidth]{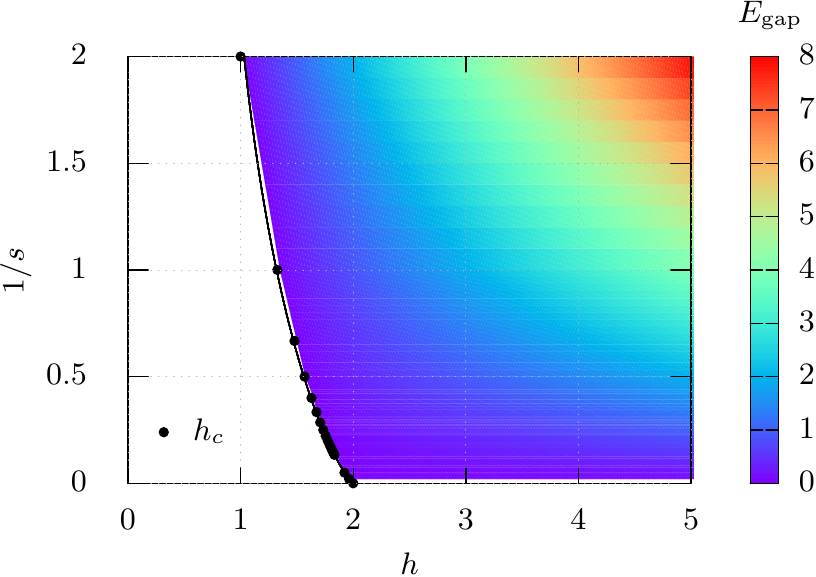}
	\caption{\label{fig:energygap}Energy gap $E_\text{gap}$ in the paramagnetic phase, as a function of the transverse field~$h$ and the inverse spin $1/s$.}
\end{figure}

\begin{figure}[b!]
	\includegraphics[width=1.0\columnwidth]{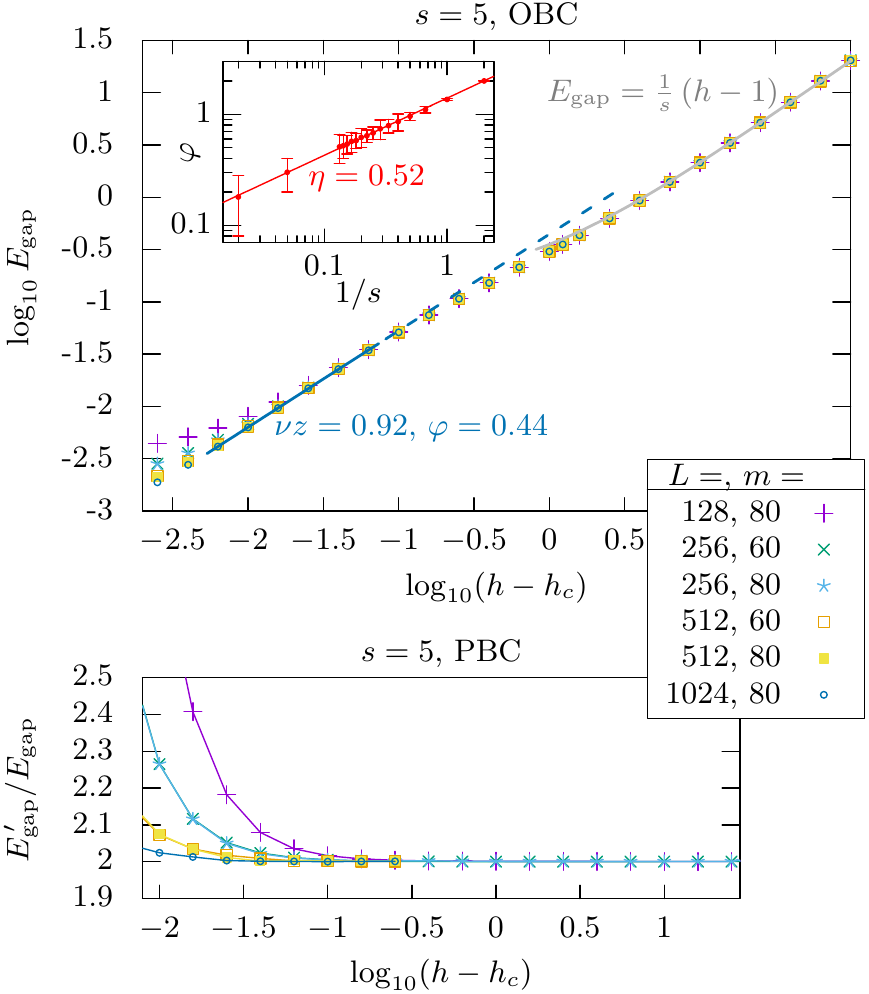}
	\caption{\label{fig:energygap-scaling}Top panel: Numerically determined energy-gap opening in the paramagnetic phase, for $s=5$ and various system sizes $L$ and bond dimensions $m$. The blue line is a power-law fit in order to determine the critical exponent $z \nu$ close to the phase transition. The inset shows the numerically determined $\varphi(s)$, together with a power-law with fitted exponent $\eta$. Bottom panel: Ratio between the accessible energy gap~$E^\prime_\text{gap}$ and the energy gap~$E_\text{gap}$, demonstrating $E^\prime_\text{gap} = 2 E_\text{gap}$, apart from finite-size effects.}
\end{figure}

We now investigate the energy gap $E_\text{gap} = E_1 - E_0$ (where $E_0$ is the ground state energy and $E_1$ is the energy of the first excited state), again as a function of both~$h$ and $s$. Fig.~\ref{fig:energygap} shows numerical data for $E_\text{gap}$ in the paramagnetic phase.
$E_\text{gap}(h)$~vanishes for $s\rightarrow\infty$, as expected for a classical model made from constituents with a continuous energy spectrum. Or, stated differently, for $s\rightarrow\infty$ excitations of arbitrarily small energy are possible because quantization vanishes. For all finite~$s$, $E_\text{gap}(h)$ scales linearly in the field strength both in immediate proximity to the phase transition, where
\begin{equation}
	E_\text{gap}(|h-h_c| \ll 1) = \varphi(s) \, |h-h_c| \; ,
\end{equation}
(using $z\nu=1$), and far from the phase transition, where
\begin{equation}
	E_\text{gap}(h \gg 1) = \frac{1}{s} \, (h - 1) \; , 
\end{equation}
as can be seen, for example, from the HP treatment outlined in Appendix~\ref{sec:analytical-hp}. We verified numerically that~$\varphi(s) > 1/s$, {\it i.e.\@} for intermediate values of~$h$ there is a transition from the steeper slope~$\varphi(s)$ to the smaller slope~$1/s$ (except for the limiting case $s=1/2$, where $\varphi(s)=1/s=\nobreak 2$). This behavior is illustrated in Fig.~\ref{fig:energygap-scaling}, using again the example $s=5$. Moreover, in the inset of Fig.~\ref{fig:energygap-scaling} we show via a fit that $\varphi(s) \approx \sqrt{2/s}$.

Finally, we note that for accurate predictions of the crossover quench time~$\tau_Q^{\times}$, the relevant quantity is the energy difference $E'_\text{gap}$ between the ground state and the lowest \textit{accessible} excited state ({\it i.e.\@} of equal parity) at the freeze-out point, which in the paramagnetic phase of the Ising model is about twice the gap~\cite{IsingDQPT}.
In the thermodynamic limit $L\to\infty$, $E'_\text{gap} = 2E_\text{gap}$ is strictly true for $s=1/2$~\cite{Lieb1961} and $s\to\infty$ (see Appendix~\ref{sec:analytical-hp}), and we verified numerically that for $h=h_\times$ it remains practically exact in all of our simulations (see lower panel of Fig.~\ref{fig:energygap-scaling} for $s=5$).
Accordingly, in Eq.~\eqref{eq:taucross} we use $\varphi=2\varphi(s)$.

\section{Comparison between DGC and ``traditional'' Ginzburg criterion} \label{sec:comparison-criteria}

\begin{figure}[b]
	\includegraphics[width=0.98\columnwidth]{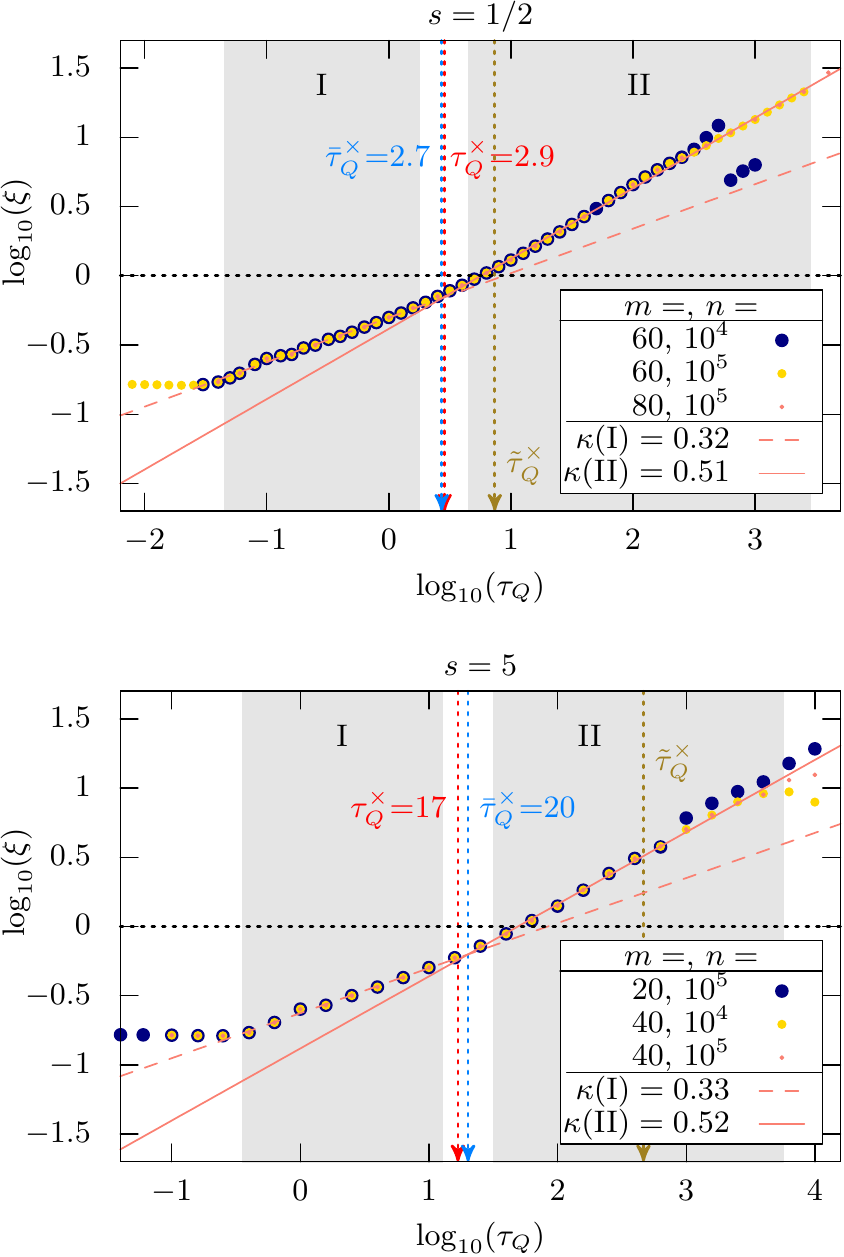}
	\caption{\label{fig:KZcompare2} Extension of Fig.~\ref{fig:KZcompare}, which includes the {\it naive} crossover timescales $\tilde{\tau}_Q^{\times}$ (brown dashed arrows), predicted by using the mean-field critical point $h'_\times = 2$ as the phase crossover point.}
\end{figure}

Here we show numerical evidence that our strategy for estimating the crossover timescale $\tau_Q^{\times}$, summarized by Eq.~\eqref{eq:taucross}, delivers better predictions than more naive approaches, at least for the class of quantum models considered here.
Specifically, in Ref.~\onlinecite{PsiKZCrossover2016} the crossover timescale $\tau_Q^{\times}$ was estimated by using the mean-field critical point $h'_\times = h^\text{MF}_c = 2$ as the phase crossover point, instead of the equilibrium point $h_\times$ where the correlation length matches the lattice spacing. In Fig.~\ref{fig:KZcompare2} we explicitly show that our estimator delivers more accurate quantitative predictions of the crossover, especially in the case of large spin $s$, where $| h_\times - h_c|$ and $|h_\times' - h_c |$ differ by orders of magnitude.

\bibliography{Ginz_refs}

\begin{thebibliography}{50}%
\makeatletter
\providecommand \@ifxundefined [1]{%
 \@ifx{#1\undefined}
}%
\providecommand \@ifnum [1]{%
 \ifnum #1\expandafter \@firstoftwo
 \else \expandafter \@secondoftwo
 \fi
}%
\providecommand \@ifx [1]{%
 \ifx #1\expandafter \@firstoftwo
 \else \expandafter \@secondoftwo
 \fi
}%
\providecommand \natexlab [1]{#1}%
\providecommand \enquote  [1]{``#1''}%
\providecommand \bibnamefont  [1]{#1}%
\providecommand \bibfnamefont [1]{#1}%
\providecommand \citenamefont [1]{#1}%
\providecommand \href@noop [0]{\@secondoftwo}%
\providecommand \href [0]{\begingroup \@sanitize@url \@href}%
\providecommand \@href[1]{\@@startlink{#1}\@@href}%
\providecommand \@@href[1]{\endgroup#1\@@endlink}%
\providecommand \@sanitize@url [0]{\catcode `\\12\catcode `\$12\catcode
  `\&12\catcode `\#12\catcode `\^12\catcode `\_12\catcode `\%12\relax}%
\providecommand \@@startlink[1]{}%
\providecommand \@@endlink[0]{}%
\providecommand \url  [0]{\begingroup\@sanitize@url \@url }%
\providecommand \@url [1]{\endgroup\@href {#1}{\urlprefix }}%
\providecommand \urlprefix  [0]{URL }%
\providecommand \Eprint [0]{\href }%
\providecommand \doibase [0]{http://dx.doi.org/}%
\providecommand \selectlanguage [0]{\@gobble}%
\providecommand \bibinfo  [0]{\@secondoftwo}%
\providecommand \bibfield  [0]{\@secondoftwo}%
\providecommand \translation [1]{[#1]}%
\providecommand \BibitemOpen [0]{}%
\providecommand \bibitemStop [0]{}%
\providecommand \bibitemNoStop [0]{.\EOS\space}%
\providecommand \EOS [0]{\spacefactor3000\relax}%
\providecommand \BibitemShut  [1]{\csname bibitem#1\endcsname}%
\let\auto@bib@innerbib\@empty
\bibitem [{\citenamefont {Kibble}(1976)}]{Kibble76}%
  \BibitemOpen
  \bibfield  {author} {\bibinfo {author} {\bibfnamefont {T.~W.~B.}\
  \bibnamefont {Kibble}},\ }\href {\doibase 10.1088/0305-4470/9/8/029}
  {\bibfield  {journal} {\bibinfo  {journal} {J. Phys. A: Math. Gen.}\ }\textbf
  {\bibinfo {volume} {9}},\ \bibinfo {pages} {1387} (\bibinfo {year}
  {1976})}\BibitemShut {NoStop}%
\bibitem [{\citenamefont {Zurek}(1985)}]{Zurek85}%
  \BibitemOpen
  \bibfield  {author} {\bibinfo {author} {\bibfnamefont {W.~H.}\ \bibnamefont
  {Zurek}},\ }\href {\doibase 10.1038/317505a0} {\bibfield  {journal} {\bibinfo
   {journal} {Nature}\ }\textbf {\bibinfo {volume} {317}},\ \bibinfo {pages}
  {505} (\bibinfo {year} {1985})}\BibitemShut {NoStop}%
\bibitem [{\citenamefont {Kibble}(1980)}]{Kibbleagain1980}%
  \BibitemOpen
  \bibfield  {author} {\bibinfo {author} {\bibfnamefont {T.~W.~B.}\
  \bibnamefont {Kibble}},\ }\href {\doibase 10.1016/0370-1573(80)90091-5}
  {\bibfield  {journal} {\bibinfo  {journal} {Phys. Rep.}\ }\textbf {\bibinfo
  {volume} {67}},\ \bibinfo {pages} {183 } (\bibinfo {year}
  {1980})}\BibitemShut {NoStop}%
\bibitem [{\citenamefont {Dziarmaga}(1998)}]{Dziarmaga98}%
  \BibitemOpen
  \bibfield  {author} {\bibinfo {author} {\bibfnamefont {J.}~\bibnamefont
  {Dziarmaga}},\ }\href {\doibase 10.1103/PhysRevLett.81.1551} {\bibfield
  {journal} {\bibinfo  {journal} {Phys. Rev. Lett.}\ }\textbf {\bibinfo
  {volume} {81}},\ \bibinfo {pages} {1551} (\bibinfo {year}
  {1998})}\BibitemShut {NoStop}%
\bibitem [{\citenamefont {Chuang}\ \emph {et~al.}(1991)\citenamefont {Chuang},
  \citenamefont {Durrer}, \citenamefont {Turok},\ and\ \citenamefont
  {Yurke}}]{Chuang91}%
  \BibitemOpen
  \bibfield  {author} {\bibinfo {author} {\bibfnamefont {I.}~\bibnamefont
  {Chuang}}, \bibinfo {author} {\bibfnamefont {R.}~\bibnamefont {Durrer}},
  \bibinfo {author} {\bibfnamefont {N.}~\bibnamefont {Turok}}, \ and\ \bibinfo
  {author} {\bibfnamefont {B.}~\bibnamefont {Yurke}},\ }\href {\doibase
  10.1126/science.251.4999.1336} {\bibfield  {journal} {\bibinfo  {journal}
  {Science}\ }\textbf {\bibinfo {volume} {251}},\ \bibinfo {pages} {1336}
  (\bibinfo {year} {1991})}\BibitemShut {NoStop}%
\bibitem [{\citenamefont {Ducci}\ \emph {et~al.}(1999)\citenamefont {Ducci},
  \citenamefont {Ramazza}, \citenamefont {Gonz{\'a}lez-Vi{\~n}as},\ and\
  \citenamefont {Arecchi}}]{Ducci99}%
  \BibitemOpen
  \bibfield  {author} {\bibinfo {author} {\bibfnamefont {S.}~\bibnamefont
  {Ducci}}, \bibinfo {author} {\bibfnamefont {P.~L.}\ \bibnamefont {Ramazza}},
  \bibinfo {author} {\bibfnamefont {W.}~\bibnamefont {Gonz{\'a}lez-Vi{\~n}as}},
  \ and\ \bibinfo {author} {\bibfnamefont {F.~T.}\ \bibnamefont {Arecchi}},\
  }\href {\doibase 10.1103/physrevlett.83.5210} {\bibfield  {journal} {\bibinfo
   {journal} {Phys. Rev. Lett.}\ }\textbf {\bibinfo {volume} {83}},\ \bibinfo
  {pages} {5210} (\bibinfo {year} {1999})}\BibitemShut {NoStop}%
\bibitem [{\citenamefont {Zurek}\ \emph {et~al.}(2005)\citenamefont {Zurek},
  \citenamefont {Dorner},\ and\ \citenamefont {Zoller}}]{IsingDQPT}%
  \BibitemOpen
  \bibfield  {author} {\bibinfo {author} {\bibfnamefont {W.~H.}\ \bibnamefont
  {Zurek}}, \bibinfo {author} {\bibfnamefont {U.}~\bibnamefont {Dorner}}, \
  and\ \bibinfo {author} {\bibfnamefont {P.}~\bibnamefont {Zoller}},\ }\href
  {\doibase 10.1103/PhysRevLett.95.105701} {\bibfield  {journal} {\bibinfo
  {journal} {Phys. Rev. Lett.}\ }\textbf {\bibinfo {volume} {95}},\ \bibinfo
  {pages} {105701} (\bibinfo {year} {2005})}\BibitemShut {NoStop}%
\bibitem [{\citenamefont {Dziarmaga}(2005)}]{IsingDQPT2}%
  \BibitemOpen
  \bibfield  {author} {\bibinfo {author} {\bibfnamefont {J.}~\bibnamefont
  {Dziarmaga}},\ }\href {\doibase 10.1103/PhysRevLett.95.245701} {\bibfield
  {journal} {\bibinfo  {journal} {Phys. Rev. Lett.}\ }\textbf {\bibinfo
  {volume} {95}},\ \bibinfo {pages} {245701} (\bibinfo {year}
  {2005})}\BibitemShut {NoStop}%
\bibitem [{\citenamefont {del Campo}\ \emph {et~al.}(2012)\citenamefont {del
  Campo}, \citenamefont {Rams},\ and\ \citenamefont {Zurek}}]{IsingDQPT3}%
  \BibitemOpen
  \bibfield  {author} {\bibinfo {author} {\bibfnamefont {A.}~\bibnamefont {del
  Campo}}, \bibinfo {author} {\bibfnamefont {M.~M.}\ \bibnamefont {Rams}}, \
  and\ \bibinfo {author} {\bibfnamefont {W.~H.}\ \bibnamefont {Zurek}},\ }\href
  {\doibase 10.1103/PhysRevLett.109.115703} {\bibfield  {journal} {\bibinfo
  {journal} {Phys. Rev. Lett.}\ }\textbf {\bibinfo {volume} {109}},\ \bibinfo
  {pages} {115703} (\bibinfo {year} {2012})}\BibitemShut {NoStop}%
\bibitem [{\citenamefont {Navon}\ \emph {et~al.}(2015)\citenamefont {Navon},
  \citenamefont {Gaunt}, \citenamefont {Smith},\ and\ \citenamefont
  {Hadzibabic}}]{hadzibacic:2015}%
  \BibitemOpen
  \bibfield  {author} {\bibinfo {author} {\bibfnamefont {N.}~\bibnamefont
  {Navon}}, \bibinfo {author} {\bibfnamefont {A.~L.}\ \bibnamefont {Gaunt}},
  \bibinfo {author} {\bibfnamefont {R.~P.}\ \bibnamefont {Smith}}, \ and\
  \bibinfo {author} {\bibfnamefont {Z.}~\bibnamefont {Hadzibabic}},\ }\href
  {\doibase 10.1126/science.1258676} {\bibfield  {journal} {\bibinfo  {journal}
  {Science}\ }\textbf {\bibinfo {volume} {347}},\ \bibinfo {pages} {167}
  (\bibinfo {year} {2015})}\BibitemShut {NoStop}%
\bibitem [{\citenamefont {Polkovnikov}(2005)}]{Polkovnikov:2005}%
  \BibitemOpen
  \bibfield  {author} {\bibinfo {author} {\bibfnamefont {A.}~\bibnamefont
  {Polkovnikov}},\ }\href {\doibase 10.1103/PhysRevB.72.161201} {\bibfield
  {journal} {\bibinfo  {journal} {Phys. Rev. B}\ }\textbf {\bibinfo {volume}
  {72}},\ \bibinfo {pages} {161201(R)} (\bibinfo {year} {2005})}\BibitemShut
  {NoStop}%
\bibitem [{\citenamefont {Chandran}\ \emph {et~al.}(2012)\citenamefont
  {Chandran}, \citenamefont {Erez}, \citenamefont {Gubser},\ and\ \citenamefont
  {Sondhi}}]{Sondhi:2012}%
  \BibitemOpen
  \bibfield  {author} {\bibinfo {author} {\bibfnamefont {A.}~\bibnamefont
  {Chandran}}, \bibinfo {author} {\bibfnamefont {A.}~\bibnamefont {Erez}},
  \bibinfo {author} {\bibfnamefont {S.~S.}\ \bibnamefont {Gubser}}, \ and\
  \bibinfo {author} {\bibfnamefont {S.~L.}\ \bibnamefont {Sondhi}},\ }\href
  {\doibase 10.1103/PhysRevB.86.064304} {\bibfield  {journal} {\bibinfo
  {journal} {Phys. Rev. B}\ }\textbf {\bibinfo {volume} {86}},\ \bibinfo
  {pages} {064304} (\bibinfo {year} {2012})}\BibitemShut {NoStop}%
\bibitem [{\citenamefont {De~Grandi}\ \emph {et~al.}(2011)\citenamefont
  {De~Grandi}, \citenamefont {Polkovnikov},\ and\ \citenamefont
  {Sandvik}}]{degrandi:2011}%
  \BibitemOpen
  \bibfield  {author} {\bibinfo {author} {\bibfnamefont {C.}~\bibnamefont
  {De~Grandi}}, \bibinfo {author} {\bibfnamefont {A.}~\bibnamefont
  {Polkovnikov}}, \ and\ \bibinfo {author} {\bibfnamefont {A.~W.}\ \bibnamefont
  {Sandvik}},\ }\href {\doibase 10.1103/PhysRevB.84.224303} {\bibfield
  {journal} {\bibinfo  {journal} {Phys. Rev. B}\ }\textbf {\bibinfo {volume}
  {84}},\ \bibinfo {pages} {224303} (\bibinfo {year} {2011})}\BibitemShut
  {NoStop}%
\bibitem [{\citenamefont {Maniv}\ \emph {et~al.}(2003)\citenamefont {Maniv},
  \citenamefont {Polturak},\ and\ \citenamefont {Koren}}]{Polturak:2003}%
  \BibitemOpen
  \bibfield  {author} {\bibinfo {author} {\bibfnamefont {A.}~\bibnamefont
  {Maniv}}, \bibinfo {author} {\bibfnamefont {E.}~\bibnamefont {Polturak}}, \
  and\ \bibinfo {author} {\bibfnamefont {G.}~\bibnamefont {Koren}},\ }\href
  {\doibase 10.1103/PhysRevLett.91.197001} {\bibfield  {journal} {\bibinfo
  {journal} {Phys. Rev. Lett.}\ }\textbf {\bibinfo {volume} {91}},\ \bibinfo
  {pages} {197001} (\bibinfo {year} {2003})}\BibitemShut {NoStop}%
\bibitem [{\citenamefont {Labeyrie}\ and\ \citenamefont
  {Kaiser}(2016)}]{KZ2016Labeyrie}%
  \BibitemOpen
  \bibfield  {author} {\bibinfo {author} {\bibfnamefont {G.}~\bibnamefont
  {Labeyrie}}\ and\ \bibinfo {author} {\bibfnamefont {R.}~\bibnamefont
  {Kaiser}},\ }\href {\doibase 10.1103/PhysRevLett.117.275701} {\bibfield
  {journal} {\bibinfo  {journal} {Phys. Rev. Lett.}\ }\textbf {\bibinfo
  {volume} {117}},\ \bibinfo {pages} {275701} (\bibinfo {year}
  {2016})}\BibitemShut {NoStop}%
\bibitem [{\citenamefont {Pal}\ \emph {et~al.}(2017)\citenamefont {Pal},
  \citenamefont {Tradonsky}, \citenamefont {Chriki}, \citenamefont {Friesem},\
  and\ \citenamefont {Davidson}}]{KZ2017Vishwa}%
  \BibitemOpen
  \bibfield  {author} {\bibinfo {author} {\bibfnamefont {V.}~\bibnamefont
  {Pal}}, \bibinfo {author} {\bibfnamefont {C.}~\bibnamefont {Tradonsky}},
  \bibinfo {author} {\bibfnamefont {R.}~\bibnamefont {Chriki}}, \bibinfo
  {author} {\bibfnamefont {A.~A.}\ \bibnamefont {Friesem}}, \ and\ \bibinfo
  {author} {\bibfnamefont {N.}~\bibnamefont {Davidson}},\ }\href {\doibase
  10.1103/PhysRevLett.119.013902} {\bibfield  {journal} {\bibinfo  {journal}
  {Phys. Rev. Lett.}\ }\textbf {\bibinfo {volume} {119}},\ \bibinfo {pages}
  {013902} (\bibinfo {year} {2017})}\BibitemShut {NoStop}%
\bibitem [{\citenamefont {Berdanier}\ \emph {et~al.}(2017)\citenamefont
  {Berdanier}, \citenamefont {Kolodrubetz}, \citenamefont {Vasseur},\ and\
  \citenamefont {Moore}}]{KZ2017Berdanier}%
  \BibitemOpen
  \bibfield  {author} {\bibinfo {author} {\bibfnamefont {W.}~\bibnamefont
  {Berdanier}}, \bibinfo {author} {\bibfnamefont {M.}~\bibnamefont
  {Kolodrubetz}}, \bibinfo {author} {\bibfnamefont {R.}~\bibnamefont
  {Vasseur}}, \ and\ \bibinfo {author} {\bibfnamefont {J.~E.}\ \bibnamefont
  {Moore}},\ }\href {\doibase 10.1103/PhysRevLett.118.260602} {\bibfield
  {journal} {\bibinfo  {journal} {Phys. Rev. Lett.}\ }\textbf {\bibinfo
  {volume} {118}},\ \bibinfo {pages} {260602} (\bibinfo {year}
  {2017})}\BibitemShut {NoStop}%
\bibitem [{\citenamefont {Meier}\ \emph {et~al.}(2017)\citenamefont {Meier},
  \citenamefont {Lilienblum}, \citenamefont {Griffin}, \citenamefont {Conder},
  \citenamefont {Pomjakushina}, \citenamefont {Yan}, \citenamefont {Bourret},
  \citenamefont {Meier}, \citenamefont {Lichtenberg}, \citenamefont {Salje},
  \citenamefont {Spaldin}, \citenamefont {Fiebig},\ and\ \citenamefont
  {Cano}}]{KZ2017Meier}%
  \BibitemOpen
  \bibfield  {author} {\bibinfo {author} {\bibfnamefont {Q.~N.}\ \bibnamefont
  {Meier}}, \bibinfo {author} {\bibfnamefont {M.}~\bibnamefont {Lilienblum}},
  \bibinfo {author} {\bibfnamefont {S.~M.}\ \bibnamefont {Griffin}}, \bibinfo
  {author} {\bibfnamefont {K.}~\bibnamefont {Conder}}, \bibinfo {author}
  {\bibfnamefont {E.}~\bibnamefont {Pomjakushina}}, \bibinfo {author}
  {\bibfnamefont {Z.}~\bibnamefont {Yan}}, \bibinfo {author} {\bibfnamefont
  {E.}~\bibnamefont {Bourret}}, \bibinfo {author} {\bibfnamefont
  {D.}~\bibnamefont {Meier}}, \bibinfo {author} {\bibfnamefont
  {F.}~\bibnamefont {Lichtenberg}}, \bibinfo {author} {\bibfnamefont
  {E.~K.~H.}\ \bibnamefont {Salje}}, \bibinfo {author} {\bibfnamefont {N.~A.}\
  \bibnamefont {Spaldin}}, \bibinfo {author} {\bibfnamefont {M.}~\bibnamefont
  {Fiebig}}, \ and\ \bibinfo {author} {\bibfnamefont {A.}~\bibnamefont
  {Cano}},\ }\href {\doibase 10.1103/PhysRevX.7.041014} {\bibfield  {journal}
  {\bibinfo  {journal} {Phys. Rev. X}\ }\textbf {\bibinfo {volume} {7}},\
  \bibinfo {pages} {041014} (\bibinfo {year} {2017})}\BibitemShut {NoStop}%
\bibitem [{\citenamefont {Kennes}\ \emph {et~al.}(2018)\citenamefont {Kennes},
  \citenamefont {de~la Torre}, \citenamefont {Ron}, \citenamefont {Hsieh},\
  and\ \citenamefont {Millis}}]{KZ2018Kennes}%
  \BibitemOpen
  \bibfield  {author} {\bibinfo {author} {\bibfnamefont {D.~M.}\ \bibnamefont
  {Kennes}}, \bibinfo {author} {\bibfnamefont {A.}~\bibnamefont {de~la Torre}},
  \bibinfo {author} {\bibfnamefont {A.}~\bibnamefont {Ron}}, \bibinfo {author}
  {\bibfnamefont {D.}~\bibnamefont {Hsieh}}, \ and\ \bibinfo {author}
  {\bibfnamefont {A.~J.}\ \bibnamefont {Millis}},\ }\href {\doibase
  10.1103/PhysRevLett.120.127601} {\bibfield  {journal} {\bibinfo  {journal}
  {Phys. Rev. Lett.}\ }\textbf {\bibinfo {volume} {120}},\ \bibinfo {pages}
  {127601} (\bibinfo {year} {2018})}\BibitemShut {NoStop}%
\bibitem [{\citenamefont {Saberi}\ \emph {et~al.}(2014)\citenamefont {Saberi},
  \citenamefont {Opatrn{\`y}}, \citenamefont {M{\o}lmer},\ and\ \citenamefont
  {del Campo}}]{Adolfo2014}%
  \BibitemOpen
  \bibfield  {author} {\bibinfo {author} {\bibfnamefont {H.}~\bibnamefont
  {Saberi}}, \bibinfo {author} {\bibfnamefont {T.}~\bibnamefont {Opatrn{\`y}}},
  \bibinfo {author} {\bibfnamefont {K.}~\bibnamefont {M{\o}lmer}}, \ and\
  \bibinfo {author} {\bibfnamefont {A.}~\bibnamefont {del Campo}},\ }\href
  {\doibase 10.1103/PhysRevA.90.060301} {\bibfield  {journal} {\bibinfo
  {journal} {Phys. Rev. A}\ }\textbf {\bibinfo {volume} {90}},\ \bibinfo
  {pages} {060301(R)} (\bibinfo {year} {2014})}\BibitemShut {NoStop}%
\bibitem [{\citenamefont {Georgescu}\ \emph {et~al.}(2014)\citenamefont
  {Georgescu}, \citenamefont {Ashhab},\ and\ \citenamefont
  {Nori}}]{Georgescu2014}%
  \BibitemOpen
  \bibfield  {author} {\bibinfo {author} {\bibfnamefont {I.~M.}\ \bibnamefont
  {Georgescu}}, \bibinfo {author} {\bibfnamefont {S.}~\bibnamefont {Ashhab}}, \
  and\ \bibinfo {author} {\bibfnamefont {F.}~\bibnamefont {Nori}},\ }\href
  {\doibase 10.1103/RevModPhys.86.153} {\bibfield  {journal} {\bibinfo
  {journal} {Rev. Mod. Phys.}\ }\textbf {\bibinfo {volume} {86}},\ \bibinfo
  {pages} {153} (\bibinfo {year} {2014})}\BibitemShut {NoStop}%
\bibitem [{\citenamefont {Lucas}(2014)}]{Lucas2014}%
  \BibitemOpen
  \bibfield  {author} {\bibinfo {author} {\bibfnamefont {A.}~\bibnamefont
  {Lucas}},\ }\href {\doibase 10.3389/fphy.2014.00005} {\bibfield  {journal}
  {\bibinfo  {journal} {Front. Phys.}\ }\textbf {\bibinfo {volume} {2}},\
  \bibinfo {pages} {5} (\bibinfo {year} {2014})}\BibitemShut {NoStop}%
\bibitem [{\citenamefont {Silvi}\ \emph {et~al.}(2016)\citenamefont {Silvi},
  \citenamefont {Morigi}, \citenamefont {Calarco},\ and\ \citenamefont
  {Montangero}}]{PsiKZCrossover2016}%
  \BibitemOpen
  \bibfield  {author} {\bibinfo {author} {\bibfnamefont {P.}~\bibnamefont
  {Silvi}}, \bibinfo {author} {\bibfnamefont {G.}~\bibnamefont {Morigi}},
  \bibinfo {author} {\bibfnamefont {T.}~\bibnamefont {Calarco}}, \ and\
  \bibinfo {author} {\bibfnamefont {S.}~\bibnamefont {Montangero}},\ }\href
  {\doibase 10.1103/PhysRevLett.116.225701} {\bibfield  {journal} {\bibinfo
  {journal} {Phys. Rev. Lett.}\ }\textbf {\bibinfo {volume} {116}},\ \bibinfo
  {pages} {225701} (\bibinfo {year} {2016})}\BibitemShut {NoStop}%
\bibitem [{\citenamefont {Amit}(1974)}]{Amit}%
  \BibitemOpen
  \bibfield  {author} {\bibinfo {author} {\bibfnamefont {D.~J.}\ \bibnamefont
  {Amit}},\ }\href {http://stacks.iop.org/0022-3719/7/i=18/a=020} {\bibfield
  {journal} {\bibinfo  {journal} {J. Phys. C: Solid State Phys.}\ }\textbf
  {\bibinfo {volume} {7}},\ \bibinfo {pages} {3369} (\bibinfo {year}
  {1974})}\BibitemShut {NoStop}%
\bibitem [{\citenamefont {Damski}(2005)}]{Damskisimplest}%
  \BibitemOpen
  \bibfield  {author} {\bibinfo {author} {\bibfnamefont {B.}~\bibnamefont
  {Damski}},\ }\href {\doibase 10.1103/PhysRevLett.95.035701} {\bibfield
  {journal} {\bibinfo  {journal} {Phys. Rev. Lett.}\ }\textbf {\bibinfo
  {volume} {95}},\ \bibinfo {pages} {035701} (\bibinfo {year}
  {2005})}\BibitemShut {NoStop}%
\bibitem [{\citenamefont {Laguna}\ and\ \citenamefont
  {Zurek}(1997)}]{Lagunaonequarter}%
  \BibitemOpen
  \bibfield  {author} {\bibinfo {author} {\bibfnamefont {P.}~\bibnamefont
  {Laguna}}\ and\ \bibinfo {author} {\bibfnamefont {W.~H.}\ \bibnamefont
  {Zurek}},\ }\href {\doibase 10.1103/physrevlett.78.2519} {\bibfield
  {journal} {\bibinfo  {journal} {Phys. Rev. Lett.}\ }\textbf {\bibinfo
  {volume} {78}},\ \bibinfo {pages} {2519} (\bibinfo {year}
  {1997})}\BibitemShut {NoStop}%
\bibitem [{\citenamefont {Laguna}\ and\ \citenamefont
  {Zurek}(1998)}]{Lagunaonethird}%
  \BibitemOpen
  \bibfield  {author} {\bibinfo {author} {\bibfnamefont {P.}~\bibnamefont
  {Laguna}}\ and\ \bibinfo {author} {\bibfnamefont {W.~H.}\ \bibnamefont
  {Zurek}},\ }\href {\doibase 10.1103/PhysRevD.58.085021} {\bibfield  {journal}
  {\bibinfo  {journal} {Phys. Rev. D}\ }\textbf {\bibinfo {volume} {58}},\
  \bibinfo {pages} {085021} (\bibinfo {year} {1998})}\BibitemShut {NoStop}%
\bibitem [{\citenamefont {White}(1992)}]{White1992}%
  \BibitemOpen
  \bibfield  {author} {\bibinfo {author} {\bibfnamefont {S.~R.}\ \bibnamefont
  {White}},\ }\href {\doibase 10.1103/PhysRevLett.69.2863} {\bibfield
  {journal} {\bibinfo  {journal} {Phys. Rev. Lett.}\ }\textbf {\bibinfo
  {volume} {69}},\ \bibinfo {pages} {2863} (\bibinfo {year}
  {1992})}\BibitemShut {NoStop}%
\bibitem [{\citenamefont {Schollw{\"{o}}ck}(2011)}]{MPSAge}%
  \BibitemOpen
  \bibfield  {author} {\bibinfo {author} {\bibfnamefont {U.}~\bibnamefont
  {Schollw{\"{o}}ck}},\ }\href {\doibase 10.1016/j.aop.2010.09.012} {\bibfield
  {journal} {\bibinfo  {journal} {Ann. Phys.}\ }\textbf {\bibinfo {volume}
  {326}},\ \bibinfo {pages} {96} (\bibinfo {year} {2011})}\BibitemShut
  {NoStop}%
\bibitem [{\citenamefont {Gerster}\ \emph {et~al.}(2014)\citenamefont
  {Gerster}, \citenamefont {Silvi}, \citenamefont {Rizzi}, \citenamefont
  {Fazio}, \citenamefont {Calarco},\ and\ \citenamefont
  {Montangero}}]{Gerster2014b}%
  \BibitemOpen
  \bibfield  {author} {\bibinfo {author} {\bibfnamefont {M.}~\bibnamefont
  {Gerster}}, \bibinfo {author} {\bibfnamefont {P.}~\bibnamefont {Silvi}},
  \bibinfo {author} {\bibfnamefont {M.}~\bibnamefont {Rizzi}}, \bibinfo
  {author} {\bibfnamefont {R.}~\bibnamefont {Fazio}}, \bibinfo {author}
  {\bibfnamefont {T.}~\bibnamefont {Calarco}}, \ and\ \bibinfo {author}
  {\bibfnamefont {S.}~\bibnamefont {Montangero}},\ }\href {\doibase
  10.1103/PhysRevB.90.125154} {\bibfield  {journal} {\bibinfo  {journal} {Phys.
  Rev. B}\ }\textbf {\bibinfo {volume} {90}},\ \bibinfo {pages} {125154}
  (\bibinfo {year} {2014})}\BibitemShut {NoStop}%
\bibitem [{\citenamefont {Vidal}(2004)}]{Vidal2004}%
  \BibitemOpen
  \bibfield  {author} {\bibinfo {author} {\bibfnamefont {G.}~\bibnamefont
  {Vidal}},\ }\href {\doibase 10.1103/physrevlett.93.040502} {\bibfield
  {journal} {\bibinfo  {journal} {Phys. Rev. Lett.}\ }\textbf {\bibinfo
  {volume} {93}},\ \bibinfo {pages} {040502} (\bibinfo {year}
  {2004})}\BibitemShut {NoStop}%
\bibitem [{\citenamefont {White}\ and\ \citenamefont
  {Feiguin}(2004)}]{TDMRGWhite}%
  \BibitemOpen
  \bibfield  {author} {\bibinfo {author} {\bibfnamefont {S.~R.}\ \bibnamefont
  {White}}\ and\ \bibinfo {author} {\bibfnamefont {A.~E.}\ \bibnamefont
  {Feiguin}},\ }\href {\doibase 10.1103/physrevlett.93.076401} {\bibfield
  {journal} {\bibinfo  {journal} {Phys. Rev. Lett.}\ }\textbf {\bibinfo
  {volume} {93}},\ \bibinfo {pages} {076401} (\bibinfo {year}
  {2004})}\BibitemShut {NoStop}%
\bibitem [{\citenamefont {Tamascelli}\ \emph {et~al.}(2015)\citenamefont
  {Tamascelli}, \citenamefont {Rosenbach},\ and\ \citenamefont
  {Plenio}}]{Tamascelli2015RSVD}%
  \BibitemOpen
  \bibfield  {author} {\bibinfo {author} {\bibfnamefont {D.}~\bibnamefont
  {Tamascelli}}, \bibinfo {author} {\bibfnamefont {R.}~\bibnamefont
  {Rosenbach}}, \ and\ \bibinfo {author} {\bibfnamefont {M.~B.}\ \bibnamefont
  {Plenio}},\ }\href {\doibase 10.1103/PhysRevE.91.063306} {\bibfield
  {journal} {\bibinfo  {journal} {Phys. Rev. E}\ }\textbf {\bibinfo {volume}
  {91}},\ \bibinfo {pages} {063306} (\bibinfo {year} {2015})}\BibitemShut
  {NoStop}%
\bibitem [{\citenamefont {Kohn}\ \emph {et~al.}(2018)\citenamefont {Kohn},
  \citenamefont {Tschirsich}, \citenamefont {Keck}, \citenamefont {Plenio},
  \citenamefont {Tamascelli},\ and\ \citenamefont {Montangero}}]{Kohn2018RSVD}%
  \BibitemOpen
  \bibfield  {author} {\bibinfo {author} {\bibfnamefont {L.}~\bibnamefont
  {Kohn}}, \bibinfo {author} {\bibfnamefont {F.}~\bibnamefont {Tschirsich}},
  \bibinfo {author} {\bibfnamefont {M.}~\bibnamefont {Keck}}, \bibinfo {author}
  {\bibfnamefont {M.~B.}\ \bibnamefont {Plenio}}, \bibinfo {author}
  {\bibfnamefont {D.}~\bibnamefont {Tamascelli}}, \ and\ \bibinfo {author}
  {\bibfnamefont {S.}~\bibnamefont {Montangero}},\ }\href {\doibase
  10.1103/PhysRevE.97.013301} {\bibfield  {journal} {\bibinfo  {journal} {Phys.
  Rev. E}\ }\textbf {\bibinfo {volume} {97}},\ \bibinfo {pages} {013301}
  (\bibinfo {year} {2018})}\BibitemShut {NoStop}%
\bibitem [{\citenamefont {Podolsky}\ \emph {et~al.}(2014)\citenamefont
  {Podolsky}, \citenamefont {Shimshoni}, \citenamefont {Silvi}, \citenamefont
  {Montangero}, \citenamefont {Calarco}, \citenamefont {Morigi},\ and\
  \citenamefont {Fishman}}]{Podolsky2014}%
  \BibitemOpen
  \bibfield  {author} {\bibinfo {author} {\bibfnamefont {D.}~\bibnamefont
  {Podolsky}}, \bibinfo {author} {\bibfnamefont {E.}~\bibnamefont {Shimshoni}},
  \bibinfo {author} {\bibfnamefont {P.}~\bibnamefont {Silvi}}, \bibinfo
  {author} {\bibfnamefont {S.}~\bibnamefont {Montangero}}, \bibinfo {author}
  {\bibfnamefont {T.}~\bibnamefont {Calarco}}, \bibinfo {author} {\bibfnamefont
  {G.}~\bibnamefont {Morigi}}, \ and\ \bibinfo {author} {\bibfnamefont
  {S.}~\bibnamefont {Fishman}},\ }\href {\doibase 10.1103/physrevb.89.214408}
  {\bibfield  {journal} {\bibinfo  {journal} {Phys. Rev. B}\ }\textbf {\bibinfo
  {volume} {89}},\ \bibinfo {pages} {214408} (\bibinfo {year}
  {2014})}\BibitemShut {NoStop}%
\bibitem [{\citenamefont {K\"uhner}\ \emph {et~al.}(2000)\citenamefont
  {K\"uhner}, \citenamefont {White},\ and\ \citenamefont
  {Monien}}]{WhiteHubbard}%
  \BibitemOpen
  \bibfield  {author} {\bibinfo {author} {\bibfnamefont {T.~D.}\ \bibnamefont
  {K\"uhner}}, \bibinfo {author} {\bibfnamefont {S.~R.}\ \bibnamefont {White}},
  \ and\ \bibinfo {author} {\bibfnamefont {H.}~\bibnamefont {Monien}},\ }\href
  {\doibase 10.1103/PhysRevB.61.12474} {\bibfield  {journal} {\bibinfo
  {journal} {Phys. Rev. B}\ }\textbf {\bibinfo {volume} {61}},\ \bibinfo
  {pages} {12474} (\bibinfo {year} {2000})}\BibitemShut {NoStop}%
\bibitem [{\citenamefont {Huang}\ \emph {et~al.}(2014)\citenamefont {Huang},
  \citenamefont {Yin}, \citenamefont {Feng},\ and\ \citenamefont
  {Zhong}}]{Huang-finite-time-scaling:2014}%
  \BibitemOpen
  \bibfield  {author} {\bibinfo {author} {\bibfnamefont {Y.}~\bibnamefont
  {Huang}}, \bibinfo {author} {\bibfnamefont {S.}~\bibnamefont {Yin}}, \bibinfo
  {author} {\bibfnamefont {B.}~\bibnamefont {Feng}}, \ and\ \bibinfo {author}
  {\bibfnamefont {F.}~\bibnamefont {Zhong}},\ }\href {\doibase
  10.1103/PhysRevB.90.134108} {\bibfield  {journal} {\bibinfo  {journal} {Phys.
  Rev. B}\ }\textbf {\bibinfo {volume} {90}},\ \bibinfo {pages} {134108}
  (\bibinfo {year} {2014})}\BibitemShut {NoStop}%
\bibitem [{\citenamefont {Liu}\ \emph {et~al.}(2014)\citenamefont {Liu},
  \citenamefont {Polkovnikov},\ and\ \citenamefont
  {Sandvik}}]{Polkovnikov-finite-time-scaling:2014}%
  \BibitemOpen
  \bibfield  {author} {\bibinfo {author} {\bibfnamefont {C.-W.}\ \bibnamefont
  {Liu}}, \bibinfo {author} {\bibfnamefont {A.}~\bibnamefont {Polkovnikov}}, \
  and\ \bibinfo {author} {\bibfnamefont {A.~W.}\ \bibnamefont {Sandvik}},\
  }\href {\doibase 10.1103/PhysRevB.89.054307} {\bibfield  {journal} {\bibinfo
  {journal} {Phys. Rev. B}\ }\textbf {\bibinfo {volume} {89}},\ \bibinfo
  {pages} {054307} (\bibinfo {year} {2014})}\BibitemShut {NoStop}%
\bibitem [{\citenamefont {Fisher}\ and\ \citenamefont
  {Barber}(1972)}]{FisherBarberFSS}%
  \BibitemOpen
  \bibfield  {author} {\bibinfo {author} {\bibfnamefont {M.~E.}\ \bibnamefont
  {Fisher}}\ and\ \bibinfo {author} {\bibfnamefont {M.~N.}\ \bibnamefont
  {Barber}},\ }\href {\doibase 10.1103/PhysRevLett.28.1516} {\bibfield
  {journal} {\bibinfo  {journal} {Phys. Rev. Lett.}\ }\textbf {\bibinfo
  {volume} {28}},\ \bibinfo {pages} {1516} (\bibinfo {year}
  {1972})}\BibitemShut {NoStop}%
\bibitem [{\citenamefont {Porras}\ and\ \citenamefont
  {Cirac}(2004)}]{PorrasCirac_AQS}%
  \BibitemOpen
  \bibfield  {author} {\bibinfo {author} {\bibfnamefont {D.}~\bibnamefont
  {Porras}}\ and\ \bibinfo {author} {\bibfnamefont {J.~I.}\ \bibnamefont
  {Cirac}},\ }\href {\doibase 10.1103/PhysRevLett.92.207901} {\bibfield
  {journal} {\bibinfo  {journal} {Phys. Rev. Lett.}\ }\textbf {\bibinfo
  {volume} {92}},\ \bibinfo {pages} {207901} (\bibinfo {year}
  {2004})}\BibitemShut {NoStop}%
\bibitem [{\citenamefont {Blatt}\ and\ \citenamefont
  {Roos}(2012)}]{RoosBlatt_AQS}%
  \BibitemOpen
  \bibfield  {author} {\bibinfo {author} {\bibfnamefont {R.}~\bibnamefont
  {Blatt}}\ and\ \bibinfo {author} {\bibfnamefont {C.~F.}\ \bibnamefont
  {Roos}},\ }\href {\doibase 10.1038/nphys2252} {\bibfield  {journal} {\bibinfo
   {journal} {Nature Physics}\ }\textbf {\bibinfo {volume} {8}},\ \bibinfo
  {pages} {277} (\bibinfo {year} {2012})}\BibitemShut {NoStop}%
\bibitem [{\citenamefont {Bernien}\ \emph {et~al.}(2017)\citenamefont
  {Bernien}, \citenamefont {Schwartz}, \citenamefont {Keesling}, \citenamefont
  {Levine}, \citenamefont {Omran}, \citenamefont {Pichler}, \citenamefont
  {Choi}, \citenamefont {Zibrov}, \citenamefont {Endres}, \citenamefont
  {Greiner}, \citenamefont {Vuleti{\'c}},\ and\ \citenamefont
  {Lukin}}]{51Rydbergs}%
  \BibitemOpen
  \bibfield  {author} {\bibinfo {author} {\bibfnamefont {H.}~\bibnamefont
  {Bernien}}, \bibinfo {author} {\bibfnamefont {S.}~\bibnamefont {Schwartz}},
  \bibinfo {author} {\bibfnamefont {A.}~\bibnamefont {Keesling}}, \bibinfo
  {author} {\bibfnamefont {H.}~\bibnamefont {Levine}}, \bibinfo {author}
  {\bibfnamefont {A.}~\bibnamefont {Omran}}, \bibinfo {author} {\bibfnamefont
  {H.}~\bibnamefont {Pichler}}, \bibinfo {author} {\bibfnamefont
  {S.}~\bibnamefont {Choi}}, \bibinfo {author} {\bibfnamefont {A.~S.}\
  \bibnamefont {Zibrov}}, \bibinfo {author} {\bibfnamefont {M.}~\bibnamefont
  {Endres}}, \bibinfo {author} {\bibfnamefont {M.}~\bibnamefont {Greiner}},
  \bibinfo {author} {\bibfnamefont {V.}~\bibnamefont {Vuleti{\'c}}}, \ and\
  \bibinfo {author} {\bibfnamefont {M.~D.}\ \bibnamefont {Lukin}},\ }\href
  {\doibase 10.1038/nature24622} {\bibfield  {journal} {\bibinfo  {journal}
  {Nature}\ }\textbf {\bibinfo {volume} {551}},\ \bibinfo {pages} {579}
  (\bibinfo {year} {2017})}\BibitemShut {NoStop}%
\bibitem [{\citenamefont {Keesling}\ \emph {et~al.}(2019)\citenamefont
  {Keesling}, \citenamefont {Omran}, \citenamefont {Levine}, \citenamefont
  {Bernien}, \citenamefont {Pichler}, \citenamefont {Choi}, \citenamefont
  {Samajdar}, \citenamefont {Schwartz}, \citenamefont {Silvi}, \citenamefont
  {Sachdev}, \citenamefont {Zoller}, \citenamefont {Endres}, \citenamefont
  {Greiner}, \citenamefont {Vuleti{\'c}},\ and\ \citenamefont
  {Lukin}}]{51RydbergsKZ}%
  \BibitemOpen
  \bibfield  {author} {\bibinfo {author} {\bibfnamefont {A.}~\bibnamefont
  {Keesling}}, \bibinfo {author} {\bibfnamefont {A.}~\bibnamefont {Omran}},
  \bibinfo {author} {\bibfnamefont {H.}~\bibnamefont {Levine}}, \bibinfo
  {author} {\bibfnamefont {H.}~\bibnamefont {Bernien}}, \bibinfo {author}
  {\bibfnamefont {H.}~\bibnamefont {Pichler}}, \bibinfo {author} {\bibfnamefont
  {S.}~\bibnamefont {Choi}}, \bibinfo {author} {\bibfnamefont {R.}~\bibnamefont
  {Samajdar}}, \bibinfo {author} {\bibfnamefont {S.}~\bibnamefont {Schwartz}},
  \bibinfo {author} {\bibfnamefont {P.}~\bibnamefont {Silvi}}, \bibinfo
  {author} {\bibfnamefont {S.}~\bibnamefont {Sachdev}}, \bibinfo {author}
  {\bibfnamefont {P.}~\bibnamefont {Zoller}}, \bibinfo {author} {\bibfnamefont
  {M.}~\bibnamefont {Endres}}, \bibinfo {author} {\bibfnamefont
  {M.}~\bibnamefont {Greiner}}, \bibinfo {author} {\bibfnamefont
  {V.}~\bibnamefont {Vuleti{\'c}}}, \ and\ \bibinfo {author} {\bibfnamefont
  {M.~D.}\ \bibnamefont {Lukin}},\ }\href {\doibase 10.1038/s41586-019-1070-1}
  {\bibfield  {journal} {\bibinfo  {journal} {Nature}\ }\textbf {\bibinfo
  {volume} {568}},\ \bibinfo {pages} {207} (\bibinfo {year}
  {2019})}\BibitemShut {NoStop}%
\bibitem [{Note1()}]{Note1}%
  \BibitemOpen
  \bibinfo {note} {{b}wUniCluster: funded by the Ministry of Science, Research
  and Arts and the universities of the state of Baden-W{\"u}rttemberg, Germany,
  within the framework program bwHPC.}\BibitemShut {Stop}%
\bibitem [{\citenamefont {Onsager}(1944)}]{Onsager44IsingModel}%
  \BibitemOpen
  \bibfield  {author} {\bibinfo {author} {\bibfnamefont {L.}~\bibnamefont
  {Onsager}},\ }\href {\doibase 10.1103/PhysRev.65.117} {\bibfield  {journal}
  {\bibinfo  {journal} {Phys. Rev.}\ }\textbf {\bibinfo {volume} {65}},\
  \bibinfo {pages} {117} (\bibinfo {year} {1944})}\BibitemShut {NoStop}%
\bibitem [{\citenamefont {Calabrese}\ and\ \citenamefont
  {Cardy}(2004)}]{Calabrese2004}%
  \BibitemOpen
  \bibfield  {author} {\bibinfo {author} {\bibfnamefont {P.}~\bibnamefont
  {Calabrese}}\ and\ \bibinfo {author} {\bibfnamefont {J.}~\bibnamefont
  {Cardy}},\ }\href {\doibase 10.1088/1742-5468/2004/06/p06002} {\bibfield
  {journal} {\bibinfo  {journal} {J. Stat. Mech: Theory Exp.}\ }\textbf
  {\bibinfo {volume} {2004}},\ \bibinfo {pages} {P06002} (\bibinfo {year}
  {2004})}\BibitemShut {NoStop}%
\bibitem [{\citenamefont {Holstein}\ and\ \citenamefont
  {Primakoff}(1940)}]{Holstein1940}%
  \BibitemOpen
  \bibfield  {author} {\bibinfo {author} {\bibfnamefont {T.}~\bibnamefont
  {Holstein}}\ and\ \bibinfo {author} {\bibfnamefont {H.}~\bibnamefont
  {Primakoff}},\ }\href {\doibase 10.1103/physrev.58.1098} {\bibfield
  {journal} {\bibinfo  {journal} {Phys. Rev.}\ }\textbf {\bibinfo {volume}
  {58}},\ \bibinfo {pages} {1098} (\bibinfo {year} {1940})}\BibitemShut
  {NoStop}%
\bibitem [{\citenamefont {Bogoliubov}(1947)}]{Bogoliubov1947}%
  \BibitemOpen
  \bibfield  {author} {\bibinfo {author} {\bibfnamefont {N.}~\bibnamefont
  {Bogoliubov}},\ }\href@noop {} {\bibfield  {journal} {\bibinfo  {journal} {J.
  Phys. (USSR)}\ }\textbf {\bibinfo {volume} {11}},\ \bibinfo {pages} {23}
  (\bibinfo {year} {1947})}\BibitemShut {NoStop}%
\bibitem [{\citenamefont {Lieb}\ \emph {et~al.}(1961)\citenamefont {Lieb},
  \citenamefont {Schultz},\ and\ \citenamefont {Mattis}}]{Lieb1961}%
  \BibitemOpen
  \bibfield  {author} {\bibinfo {author} {\bibfnamefont {E.}~\bibnamefont
  {Lieb}}, \bibinfo {author} {\bibfnamefont {T.}~\bibnamefont {Schultz}}, \
  and\ \bibinfo {author} {\bibfnamefont {D.}~\bibnamefont {Mattis}},\ }\href
  {\doibase 10.1016/0003-4916(61)90115-4} {\bibfield  {journal} {\bibinfo
  {journal} {Ann. Phys.}\ }\textbf {\bibinfo {volume} {16}},\ \bibinfo {pages}
  {407} (\bibinfo {year} {1961})}\BibitemShut {NoStop}%
\bibitem [{\citenamefont {Silvi}\ \emph {et~al.}(2019)\citenamefont {Silvi},
  \citenamefont {Tschirsich}, \citenamefont {Gerster}, \citenamefont
  {J\"unemann}, \citenamefont {Jaschke}, \citenamefont {Rizzi},\ and\
  \citenamefont {Montangero}}]{TNAnthology}%
  \BibitemOpen
  \bibfield  {author} {\bibinfo {author} {\bibfnamefont {P.}~\bibnamefont
  {Silvi}}, \bibinfo {author} {\bibfnamefont {F.}~\bibnamefont {Tschirsich}},
  \bibinfo {author} {\bibfnamefont {M.}~\bibnamefont {Gerster}}, \bibinfo
  {author} {\bibfnamefont {J.}~\bibnamefont {J\"unemann}}, \bibinfo {author}
  {\bibfnamefont {D.}~\bibnamefont {Jaschke}}, \bibinfo {author} {\bibfnamefont
  {M.}~\bibnamefont {Rizzi}}, \ and\ \bibinfo {author} {\bibfnamefont
  {S.}~\bibnamefont {Montangero}},\ }\href {\doibase
  10.21468/SciPostPhysLectNotes.8} {\bibfield  {journal} {\bibinfo  {journal}
  {SciPost Phys. Lect. Notes}\ ,\ \bibinfo {pages} {8}} (\bibinfo {year}
  {2019})}\BibitemShut {NoStop}%
\end{thebibliography}%

\end{document}